\documentclass[aps,pra,twocolumn,showpacs,nofootinbib,longbibliography,notitlepage]{revtex4-1}
\usepackage{etex}
\usepackage{amsmath,amssymb,amsthm}
\usepackage[colorlinks=true,citecolor=blue,urlcolor=blue]{hyperref}
\usepackage[pdftex]{graphicx}
\usepackage{times,txfonts}
\usepackage{braket}
\usepackage{color}
\usepackage{natbib}
\usepackage{amsmath,blkarray}
\usepackage{mathtools}
\usepackage{latexsym}
\usepackage{tabularx, booktabs}
\usepackage{graphics,epstopdf}
\usepackage{graphicx}
\usepackage{float}
\usepackage{graphicx}
\usepackage{amsfonts}
\usepackage{subcaption}
\usepackage{color,soul}

\newcommand{\be}{\begin{equation}}
	\newcommand{\ee}{\end{equation}}
\newcommand{\ba}{\begin{eqnarray}}
	\newcommand{\ea}{\end{eqnarray}}

\begin{document}
	\title{Semi-device independent certification of multiple unsharpness parameters through sequential measurements }
	\author{ Sumit Mukherjee }
	\email{mukherjeesumit93@gmail.com}
	\author{ A. K. Pan }
	\email{akp@nitp.ac.in}
	\affiliation{National Institute of Technology Patna, Ashok Rajpath, Patna, Bihar 800005, India}
	\begin{abstract}
		Based on a sequential communication game, semi-device independent certification of an unsharp instrument has recently been demonstrated [\href{https://iopscience.iop.org/article/10.1088/1367-2630/ab3773}{New J. Phys. 21 083034 (2019), }\href{https://journals.aps.org/prresearch/abstract/10.1103/PhysRevResearch.2.033014}{ Phys. Rev. Research 2, 033014 (2020)}]. In this paper, we provide semi-device independent self-testing protocols in the prepare-measure scenario to certify multiple unsharpness parameters along with the states and the measurement settings. This is achieved through the sequential quantum advantage shared by multiple independent observers in a suitable communication game known as parity-oblivious random-access-code.  We demonstrate that in 3-bit parity-oblivious random-access-code, at most three independent observers can sequentially share quantum advantage. The optimal pair (triple) of quantum advantages enables us to uniquely certify the qubit states, the measurement settings, and the unsharpness parameter(s). The practical implementation of a given protocol involves inevitable losses. In a sub-optimal scenario, we derive a certified interval within which a specific unsharpness parameter has to be confined. We extend our treatment to the 4-bit case and show that at most two observers can share quantum advantage for the qubit system. Further, we provide a sketch to argue that four sequential observers can share the quantum advantage for the two-qubit system, thereby enabling the certification of three unsharpness parameters.

	\end{abstract}
	\maketitle
	\section{Introduction}
	
	Bell's theorem \cite{bell} is regarded by many as one of the most profound discoveries of modern science. This celebrated no-go proof asserts that any ontological model satisfying locality cannot account for all quantum statistics. This feature is commonly demonstrated through the quantum violation of suitable Bell inequalities. Besides the immense conceptual insights Bell's theorem adds to the quantum foundations research, it provides a multitude of practical applications in quantum information processing (see, for extensive reviews, \cite{hororev,guhnearev,brunnerrev}). Moreover, the non-local correlation is device-independent, i.e., no characterization of the devices is needed to be assumed. Only the observed output statistics are enough to certify non-locality. The device-independent non-local correlation is used as a resource for secure quantum key distribution \cite{bar05,acin06,acin07,pir09}, randomness certification \cite{col06,pir10,nieto,col12}, witnessing Hilbert space dimension \cite{wehner,gallego,ahrens,brunnerprl13,bowler,sik16prl,cong17,pan2020} and for achieving advantages in communication complexity tasks \cite{complx1}.
	
	Optimal quantum value of a given Bell expression enables device-independent certification of state and measurements - commonly known as self-testing \cite{self1,wagner2020,tava2018,farkas2019}. For example, optimal violation of CHSH inequality self-tests the maximally entangled state and mutually anti-commuting local observables. Note that the device-independent certification encounters practical challenges arises from the requirement of the loophole-free Bell test.  Such tests have recently been realized \cite{lf1,lf2,lf3,lf4,lf5} enabling experimental demonstrations of device-independent certification of randomness \cite{Yang, Peter}. However, practically such a loophole-free certification of non-local correlation remains an uphill task.  
	
	In recent times, there is an upsurge of interest in semi-device independent (SDI)  protocols in prepare-measure scenario \cite{pawlowski11, lungi,li11,li12,Wen,van, brask, bowles14,tava2018,farkas2019,tava20exp,mir2019,wagner2020,mohan2019,miklin2020,samina2020,fole2020,anwar2020}. It is less cumbersome than full device-independent test and hence experimentally more appealing. In most of the SDI certification protocols, it is commonly assumed that there is an upper bound on dimension, but otherwise the devices remain uncharacterized. Of late, a flurry of SDI protocols has been developed for certifying the states and sharp measurements \cite{tava2018}, non-projective measurements \cite{mir2019,samina2020}, mutually unbiased bases \cite{farkas2019}, randomness  \cite{lungi,li11,li12}. Very recently, the SDI certification of unsharpness parameter has been reported \cite{miklin2020,mohan2019,anwar2020,fole2020}.

	In this work, we provide interesting SDI protocols in the prepare-measure scenario to certify multiple unsharpness parameters. This is demonstrated by examining the quantum advantage shared by multiple observers. Sharing various forms of quantum correlations by multiple sequential observers has recently received considerable attention among researchers. Quite many works \cite{silva2015,sasmal2018,bera2018,kumari2019,brown2020,Zhang21} have recently been reported to examine at most how many sequential observers can share different quantum correlations, viz., entanglement, coherence, non-locality, and preparation contextuality. Any such sharing of quantum correlation protocol requires the prior observers to perform unsharp measurements \cite{busch} represented by a set of positive operator valued measures (POVMs). This allows the subsequent observer to extract the quantum advantage. In ideal sharp measurements \cite{vonneumann}, the system collapses to one of the eigenstates of the measured observable. From an information-theoretic perspective, the information gained in sharp measurement is maximum; thereby, the system is most disturbed. It may be natural to expect that the projective measurement provides an optimal advantage in contrast to POVMs, but there are certain tasks where unsharp measurement showcases its supremacy over projective measurement. Examples include sequential quantum state discrimination \cite{std1,std2}, unbounded randomness certification \cite{ran1} and many more.

	However, the statistics corresponding to POVMs can be simulated by projective measurements unless they are extremal ones \cite{oszmaniec2017}. The unsharp POVMs that are the noisy variants of sharp projective measurements can be simulated through the classical post-processing of ideal projective measurement statistics. The certification of such unsharp non-extremal POVMs using the standard self-testing scheme is not possible. Note that the quantum measurement instruments are subject to imperfections for various reasons, and emerging quantum technologies demand certified instruments for conclusive experimental tests. Standard self-testing protocols certify the states and measurements based on the optimal quantum value, but such protocols do not certify the post-measurement states. The same optimal quantum value of a Bell expression may be obtained even when the POVMs are implemented differently, leading to different post-measurements states. But the sequential measurements have the potential to certify the post-measurement states and consequently the unsharpness parameter. 
	

	Recently, in an interesting work, such a certification was put forwarded by Mohan, Tavakoli and Brunner (henceforth, MTB) \cite{mohan2019}  through the sequential quantum advantages of two independent observers in an SDI prepare-measure communication game. In particular, they have considered $2$-bit Random-access code(RAC) and showed that two sequential observers can get the quantum advantage at most.  Both the observers cannot have optimal quantum advantages, but there exists an optimal trade-off between quantum advantages of two sequential observers, {enabling the certification of the prepared qubit states and measurement settings.}  Another interesting protocol for the same purpose is proposed in \cite{miklin2020}, but significantly differs from the approach used in  \cite{mohan2019}. MTB protocol \cite{mohan2019} also provided a certified interval of values of sharpness parameter in loss tolerant scenario. MTB proposal has been experimentally tested in \cite{anwar2020,fole2020,xiao}. 
	
	Against this backdrop, natural questions may arise, such as whether two or more independent unsharpness parameters can be certified and the certified interval of values of unsharpness parameter of a measurement instrument can be fine-tuned. Here we answer both the questions affirmatively. Intuitively, two or more unsharpness parameters can be certified if multiple independent observers can share the quantum advantage. In such a case, the certified interval of unsharpness parameters of a measurement instrument will also be fine-tuned. By using a straightforward mathematical approach, we provide SDI prepare-measure protocols to demonstrate that more than two sequential observers can share the quantum advantage, thereby enabling us to certify multiple unsharpness parameters instead of a single unsharpness parameter as it is in \cite{mohan2019,miklin2020}. However, in the practical scenario, the perfect optimal quantum correlations cannot be achieved. In such a case, we derive an interval of unsharpness parameter using the suboptimal witness pair. The closer the observed statistics are to the optimal values, the narrower is the interval to which the sharpness parameter can be confined.  We also provide a  methodology for how sequential measurement schemes can be used to fine-tune a certified interval of the unsharpness parameter of a quantum measurement device. Furthermore, we argue that the higher the number of observers shares the quantum advantage narrower the certified interval.
	
	We first propose the aforementioned SDI certification scheme based on $3$-bit RAC in a prepare-measure scenario and demonstrate that at most, three independent observers can share the quantum advantage sequentially. Throughout this paper, we assume that the quantum system to be a qubit; otherwise, the devices remain uncharacterized. We show that the optimal pair of quantum advantage corresponding to the first two observers certifies the first observer's unsharpness parameter, and the optimal triple of quantum advantage provides the certification of unsharpness parameters of the first two observers. As indicated earlier, we further provide a fine-tuning of the interval of values of unsharpness parameter compared to the $2$-bit RAC presented in \cite{mohan2019}. We also show that if the quantum advantage is extended to a third sequential observer, then the interval of values of unsharpness parameter for the first observer can be more fine-tuned, thereby certifying a narrow interval in the sub-optimal scenario. This scheme is further extended to the $4$-bit case, where we demonstrate that at least two observers can share quantum advantage for a qubit system. If the two-qubit system is taken, four independent observers can share the quantum advantage at most. We provide a sketch of how to certify three unsharpness parameters corresponding to the first three observers and the input two-qubit states and measurement settings.
	
	This paper is organized as follows. In Sec. II, we discuss a general $n$-bit quantum parity-oblivious RAC. In section III, we explicitly consider the $3$-bit case and find the optimal and sub-optimal relationships between sequential success probabilities for two observers. In Sec. IV, we generalize the scenario to three or more observers. In Sec. V we discuss the certification of multiple unsharpness parameters when the third observer gets the quantum advantage. In Sec. VI, we extend the protocol for the 4-bit case. We discuss our results in Sec. VII.
	\section{ Parity-oblivious random-access-code}
	
	We start by briefly encapsulating the notion of parity-oblivious RAC which is used here as a tool to demonstrate our results. It is a two-party one-way communication game. The $n$-bit RAC \cite{ambainis} in terms of a prepare-measure scenario that involves a sender (Alice), who has a  length-$n$ string  $x$, randomly sampled from $x\in\{0,1\}^{n}$. On the other hand, a receiver (Bob) receives uniformly at random, index $ y \in \{1,2, ..., n\}$  as his input. Bob's task is to recover the bit $x_{y}$ with a probability. In an operational theory, Alice encodes her $ n $-bit string $ x $ into the states prepared by a procedure $P_{x}$. After receiving the system, for every $y \in \{1,2,...,n\}$, Bob performs a two-outcome measurement $M_{y}$ and reports outcome $b$ as his output. As mentioned, the winning condition of the game is $b=x_{y}$, i.e., Bob has to predict the $y^{th}$ bit of Alice's input string $x$ correctly. Then the average success probability of the game is given by,
	\begin{equation}
		\label{qprob}
		S_{n}(b=x_{y}) = \dfrac{1}{2^n n}\sum\limits_{x,y}p(b=x_y|P_{x},M_{y}).
	\end{equation}
	
	Now, to help Bob, Alice can communicate some bits to him over a classical communication channel. The game becomes trivial if Alice communicates $n$ number of bits to Bob, and Non-triviality may arise if Alice communicates less than that. For our purpose here, we impose a constraint - the parity-oblivious condition \cite{spekk09} that has to be satisfied by Alice's inputs.  This demands that Alice can communicate any number of bits to Bob, but that must not reveal any information about the parity of Alice's input. The parity-oblivious condition eventually provides an upper bound on the number of bits that can be communicated.
	
	Following Spekkens \emph{et al.} \cite{spekk09}, we define a parity set $ \mathbb{P}_n= \{x|x \in \{0,1\}^n,\sum_{r} x_{r} \geq 2\} $ with $r\in \{1,2,...,n\}$. Explicitly, the parity-oblivious condition dictates that for any $s \in \mathbb{P}_{n}$, no information about $s\cdot x = \oplus_{r} s_{r}x_{r}$ (s-parity) is to be transmitted to Bob, where $\oplus$ is sum modulo $ 2 $. We then have s-parity 0 and s parity-1 sets. Explicitly, the parity-oblivious condition demands that in an operational theory the following relation is satisfied,
	
	\begin{align}
		\label{poc1}
		\forall s: \frac{1}{2^{n-1}}\sum\limits_{x|x.s=0} P(P_{x}|b,M_{y})=\frac{1}{2^{n-1}}\sum\limits_{x|x.s=1} P(P_{x}|b,M_{y}).
	\end{align}

	For example, when $ n=2 $ the set is $\mathbb{P}=\{11\} $, so no information about $x_1\oplus x_2$ can be transmitted by Alice. It was shown in  \cite{spekk09} that to satisfy the parity-oblivious condition in  the $n$-bit case, the communication from Alice to Bob has to be restricted to be one bit.

	The maximum average success probability in such a classical $ n $-bit parity-oblivious RAC is ${(n+1)}/{2n}$.  While the explicit proof can be found in \cite{spekk09}, a simple trick can saturate the bound as follows. Assume that Alice always encodes the first bit (prior agreement between Alice and Bob) and sends it to Bob.  If $y=1$, occurring with probability $ 1/n $, Bob can predict the outcome with certainty, and for $y \neq 1$, occurring with the probability of $(n-1)/n$, he at best guesses the bit with probability $1/2$. Hence the total probability of success is derived \cite{spekk09} as

	\begin{align}
		\label{cb}
		(S_{n}(b=x_{y}))_{c}\leq \frac{1}{n} + \frac{(n-1)}{2n} = \frac{1}{2}\left(1+\frac{1}{n}\right).
	\end{align}
	
	In quantum RAC, Alice encodes her length-$ n $ string $ x \in\{0,1\}^{n} $ into quantum states $ \rho_{x}^{0}$, prepared by a procedure $P_{x}$, Bob performs a two-outcome measurement $M_{y}$ for every $y \in \{1,2,...,n\}$ and reports outcome $b$ as his output. Average success probability in quantum theory can then be written as,
	\begin{eqnarray}
		\label{succ1}
		S_{n}=  \dfrac{1}{2^n n} \sum_{y=1}^{n} \sum\limits_{x \in \{0,1\}^{n}} Tr[\rho_{x}^{0} M_{b|y}].
	\end{eqnarray}  
	
	The parity-oblivious constraint {imposes} the following condition to be satisfied by Alice's input states,
	\begin{align}
		\label{poc}
		\forall s: \frac{1}{2^{n-1}}\sum\limits_{x|x.s=0} \rho_{x}^{0}=\frac{1}{2^{n-1}}\sum\limits_{x|x.s=1} \rho_{x}^{0}.
	\end{align}
	
	It has been shown in \cite{spekk09,ghorai18} that the optimal quantum success probability for $2$-bit parity-oblivious RAC is $(S_{2})^{opt}=(1/2)(1+1/\sqrt{2}) $  and  $3-$bit case is $(S_{3})^{opt}=(1/2)(1+1/\sqrt{3}) $. In both the cases, the qubit system is enough to obtain the optimal quantum value. But, for $n>3$ one requires higher dimensional system.  {In entanglement  assisted variant of parity-oblivious RAC with $n>3$, optimal values of success probabilities achievable with qubit systems were found in \cite{pan2020}.} However, in this work we consider the prepare-measure scenario and hence we separately prove the results for $n=3$ and $n=4$. 
	A remark on the ontological model of quantum theory could be useful to understand the type of non-classicality appearing in parity-oblivious quantum RAC. The satisfaction of parity-oblivious condition in an operational theory implies that no measurement can distinguish the parity of the inputs. This is regarded as an equivalent class of preparations \cite{spek05,kunjwal,pan19} which will have equivalent representation at the level of the ontic states. It has been demonstrated in \cite{spekk09} that the parity-obliviousness at the operational level must be satisfied at the level of ontic states if the ontological model of quantum theory is preparation non-contextual. Thus, in a preparation non-contextual model, the classical bound remains the same as given in Eq.(\ref{cb}). Quantum violation of this bound thus demonstrates a form of non-classicality - the preparation contextuality. Throughout this paper, by quantum advantage, we refer to the violation of preparation non-contextuality (unless stated otherwise) but to avoid clumsiness, we skip detailed discussion about it. We refer \cite{spek05,spekk09,kunjwal,pan19} for detailed discussion about it.
	
	We first consider the $3$-bit parity-oblivious RAC and demonstrate that three independent Bobs can sequentially share the quantum advantage at most by assuming the quantum system to be a qubit. We then demonstrate how this sequential measurement scenario enables the SDI certification of multiple unsharpness parameters along with the qubit states and the measurement settings. { We note here that the MTB protocol was extended \cite{wei21} for $3$-bit standard RAC  to certify a single unsharpness parameter. In our work, we consider the parity-oblivious RAC and certify multiple unsharpness parameters. While the mathematical approach in \cite{wei21} follows the  MTB protocol\cite{mohan2019}, our approach is quite different from \cite{mohan2019,wei21}. Also, we extend our approach to $4$-bit standard and parity-oblivious RACs and provide some important results in Sec. VI.}

	\section{Sequential quantum advantage in $3$-bit RAC}
	
	In sequential quantum RAC \cite{tava2018}, multiple independent observers perform a prepare-transform-measure task. Alice encodes her bit into a quantum system and sends it to the next observer (say, Bob). There are one Alice and multiple numbers of independent Bobs. After receiving the system, first Bob (say, Bob$_{1}$) performs random measurements to decode the information and relays the system to the second Bob (Bob$_{2}$), who does the same as the first Bob. The chain continues until $k^{th}$ Bob (where $k$ is arbitrary and has to be determined) gets the quantum advantage. 
	
	Before examining how many independent Bobs can get quantum advantage sequentially in a $3$-bit RAC, let us first discuss the classical RAC task in sequence. When the physical devices are classical, the states that are being sent by Alice are diagonal in the same basis on which multiple independent Bobs perform their measurements sequentially. In that case, the measurement of first Bob does not disturb the system, and hence decoding measurement for second Bob will not be influenced by the first Bob.	In other words, there exists no trade-off between sequential Bobs, and each of them can obtain optimal classical success probability. The range in which classical success probabilities of standard RAC lie is { $\frac{1}{2} \leq (S_{3}^{1},S_{3}^{2},...., S_{3}^{k}) \leq \frac{3}{4}$. However, in parity-oblivious RAC, as explained in Sec. II classical success probabilities lie in $\frac{1}{2} \leq (S_{3}^{1},S_{3}^{2},...., S_{3}^{k}) \leq \frac{2}{3}$.} In $3$-bit RAC, we will find that the input states for which the optimal quantum success probability is obtained automatically satisfy the parity-oblivious conditions. Hence, we keep our discussion in parity-oblivious RAC in the $3$-bit case.  
	
	On the other hand, in the quantum case, the prior measurements, in general, disturb Alice's input quantum states and thereby influence the success probability of the subsequent Bobs. Then there exists an information disturbance trade-off - the more information one gains from the system, the more disturbance is caused to the system. Sharp projective valued measurements disturb the system most. Then to get the quantum advantage for $k^{th}$ Bob (where $k$ is arbitrary),  previous Bobs have to measure unsharp POVMs. To obtain the maximum number of independent Bobs who can share quantum advantage, the measurements of previous Bobs have to be so unsharp that they are just enough to reveal the quantum advantage. We find that at most, two independent Bobs can share quantum advantage in $3$-bit standard RAC, but at most three Bobs can share quantum advantage in $3$-bit parity-oblivious RAC.

\begin{figure*}
		\centering
		\includegraphics[scale=0.4]{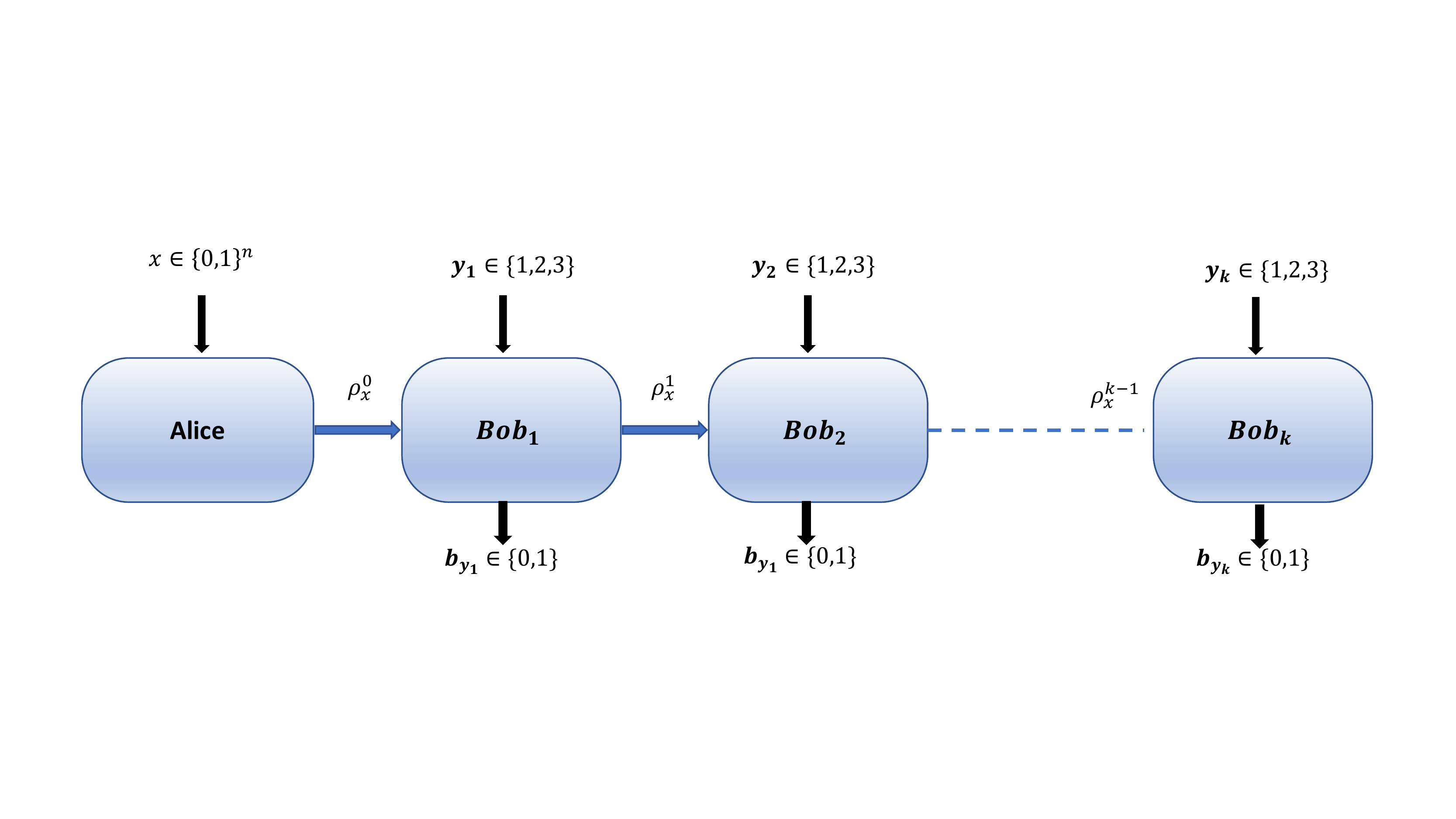}
		\caption{Prepare-transform-measure Block diagram for sequential encoding decoding scheme involving four parties. The first party gets a random three-bit string which she encodes into a qubit system. It is being sent to the next party, who thereby performs three binary outcome measurements at random to decode the message and send the system to the next party and so on}\label{fig1}
	\end{figure*}

	
	{In $3$-bit sequential quantum RAC, Alice randomly encodes her three bit string $x=(x_{0}, x_{1}, x_{2}) \in \{0,1\}^3$ into eight qubits $\rho_x^{0}$ and sends them to Bob$_{1}$. After receiving the system, Bob$_{1}$ randomly performs unsharp measurement (with unsharpness parameter $\lambda_{1}$) of dichotomic observables $B_{y_{1}}$ ( y$_{1}$=1,2,3) and relays the system to Bob$_{2}$ who carries out unsharp measurement of the observables $B_{y_{2}}$ ($y_{2}=1,2,3$) with unsharpness parameter $\lambda_{2}$ and so on. The scheme is depicted in figure Fig\ref{fig1}.  Each Bob examines if quantum advantage over classical RAC is obtained and for the $k^{th}$ Bob if the advantage is not obtained then the process stops. Let us also define the measurement observables of Bob$_{k}$ as $B_{y_{k}}$ with y$_{k}\in\{1,2,3\}$. Here we assume that a particular observer (say Bob$_{k}$) is provided with different possible measurement instruments with same value of unsharpness parameters $\lambda_{k}$.  Although, each Bob can take different values to the unsharpness parameters for different measurement settings, for our purpose it is sufficient to assume the same value of unsharpness parameters for all the measurement settings that a particular Bob uses.}
	
	If $k^{th}$ Bob's instruments are characterized by Kraus operators $\{K_{b_{k}|y_{k}}\}$ then after $k^{th}$ Bob's measurement, the average state  relayed to Bob$_{k+1}$ is given by
	\begin{equation}
		\rho^{k}_x  = \frac{1}{3}\sum_{y_{k}, b_{k}} K_{b_{k}|y_{k}}\rho_{x}^{k-1}K_{b_{k}|y_{k}}, \label{eq:D}
	\end{equation}
	where $\forall k$, $b_{k}\in \{0,1\}$  and  $\sum_{b_{k}} K_{b_{k}|y_{k}}^{\dagger}K_{b_{k}|y_{k}}=\mathbb{I}$ and {$K_{b_{k}|y_{k}}= U\sqrt{M_{b_{k}|y_{k}} }$ where $U$ is an unitary operator. For simplicity, here we set $U=\mathbb{I}$, and our arguments remains valid up to any unitary transformation.} Here $M_{b_{k}|y_{k}} = \left(\mathbb{I}+\lambda_{k}(-1)^{x_{y_{k}}} \hat{b}_{k}\cdot\sigma\right)/2$ is the POVM corresponding to the result $ b_{k}$ of measurement $B_{y_{k}}$.
	
	The Kraus operators for k$^{th}$ Bob can be written as,	$K_{\pm|y_{k}}= \sqrt{\frac{(1\pm\lambda_{k})}{2}}P_{y_{k}}^{+} +\sqrt{\frac{(1\mp\lambda_{k})}{2}}P_{y_{k}}^{-} = \alpha_{k} \mathbb{I} \pm \beta_{k} B_{y_{k}} $ where $P_{y_{k}}^{\pm}=(\mathbb{I}\pm B_{y_{k}})/2$ are the projectors and $\lambda_{k}$ is the unsharpness parameter. Here $\alpha_{k} = \frac{1}{2}\Big(\sqrt{\frac{(1-\lambda_{k})}{2}}+ \sqrt{\frac{(1+\lambda_{k})}{2}}\Big)$ and $\beta_{k} = \frac{1}{2}\Big(\sqrt{\frac{(1+\lambda_{k})}{2}}-\sqrt{\frac{(1-\lambda_{k})}{2}}\Big)$, with $\alpha_{k}^2 +\beta_{k}^2 = 1/2$ and $4\alpha_{k}\beta_{k}=\lambda_{k}$.
	
	By using Eq. (\ref{succ1}) the quantum success probability for the k$^{th}$ Bob can be written as,
	\begin{equation}
		S_{3}^{k} =\frac{1}{24}\sum_{x \in \{0,1\}^3}\sum_{y_{k}=1,2,3}Tr\Big[\rho_{x}^{k-1} M_{b_{k}|y_{k}}\Big]. \label{eq:S}
	\end{equation}
	Let us consider that Alice's eight input qubit states $\rho_{x}^{0} = \frac{\mathbb{I}+\vec{a}_{x}\cdot \sigma}{2} $ where $\vec{a}_{x}$ is the Bloch vector with $||\vec{a}_{x}||\leq 1$. By using Eq. \eqref{eq:S}, the quantum success probability for Bob$_{1}$ can be written as 
	
	\begin{equation}
		S_{3}^{1} =\frac{1}{24}\sum_{x \in \{0,1\}^3}\sum_{y_{1}=1,2,3}Tr\Big[\rho_x^{0} M_{b_{1}|y_{1}}\Big].	\label{s31}
	\end{equation}

	We consider that $ M_{b_{1}|y_{1}}$ are unbiased POVMs represented by $ M_{b_{1}|y_{1}}=(\mathbb{I}+ 4\alpha_{1}\beta_{1}(-1)^{x_{y_1}} B_{y_{1}})/2$. Using it in Eq. (\ref{s31})  and further simplifying, we get
	\begin{equation}
		S_{3}^{1}=\frac{1}{2}+\frac{\alpha_{1} \beta_{1}}{12}\sum_{r,y_{1}=1,2,3}\big(\delta_{r,y_{1}}\vec{m}_{r}\cdot \hat{b}_{y_{1}}\big), \label{eq:succ1}
	\end{equation}\\
	
	where $\vec{m}_{r}=||\vec{m}_{r}|| \hat{m}_{r}$ ($ \hat{m}_{r}$ is the unit vector) are unnormalized vectors can explicitly be written as,
	\begin{align}
		\vec{m}_1=(\vec{a}_{000}-\vec{a}_{111})+(\vec{a}_{001}-\vec{a}_{110})+(\vec{a}_{010}-\vec{a}_{101})-(\vec{a}_{100}-\vec{a}_{011}) \nonumber \\
		\vec{m}_2= (\vec{a}_{000}-\vec{a}_{111})+(\vec{a}_{001}-\vec{a}_{110})-(\vec{a}_{010}-\vec{a}_{101})+(\vec{a}_{100}-\vec{a}_{011}) \nonumber\\ 
		\vec{m}_3= (\vec{a}_{000}-\vec{a}_{111})-(\vec{a}_{001}-\vec{a}_{110})+(\vec{a}_{010}-\vec{a}_{101})+(\vec{a}_{100}-\vec{a}_{011}). \label{eq:effd}
	\end{align}
	
	Next, if Bob$_{1}$ performs unsharp measurements then the average state that is relayed to Bob$_{2}$ is obtained from Eq. \eqref{eq:D} and  is given by,
	\begin{equation}
		\rho_{x}^{1} =\frac{1}{2}\left( \mathbb{I} + \vec{a}^{1}_{x}\cdot\sigma\right), 
	\end{equation}
	where 
	\begin{equation}
		\vec{a}^{1}_{x}= 2\left(\alpha_{1}^2-\beta_{1}^{2}\right)\vec{a}_{x} +\frac{4\beta_{1}^2}{3}\sum_{y_{1}=1,2,3}\left( \hat{b}_{y_{1}}\cdot\vec{a}_{x}\right) \hat{b}_{y_{1}} \label{eq:vecA}
	\end{equation}
	is the Bloch vector of the reduced state. The quantum success probability for Bob$_{2}$ is calculated by using Eqs.\eqref{eq:S} and \eqref{eq:vecA} as,
	\begin{equation}
		\begin{split}
			S_{3}^{2} &=\frac{1}{24}\sum_{x \in \{0,1\}^3}\sum_{y_{2}=1,2,3}Tr\Big[\rho_{x}^1 M_{b_{2}|y_{2}}\Big]\\
			&= \frac{1}{2}+\frac{1}{48}\Big[2\left(\alpha_{1}^2-\beta_{1}^{2}\right)\sum_{r,y_{2}=1,2,3}\big(\delta_{r,y_{2}}\vec{m}_{r}\cdot \hat{b}_{y_{2}}\big)\\
			&+\frac{4\beta_{1}^2}{3} \sum_{r,y_{1}, y_{2}=1,2,3}\left(\delta_{r,y_{2}}\vec{m}_{r}\cdot \hat{b}_{y_{1}}\right)\left( \hat{b}_{y_{1}}\cdot \hat{b}_{y_{2}}\right)\Big]. \label{eq:succ2}
		\end{split}
	\end{equation}
	
	In order to find the optimal trade-off between the success probabilities of Bob$_{1}$ and Bob$_{2}$, the maximum quantum value of $S_{3}^{2}$ for the given value of $S_{3}^{1}$ has to be derived. This leads {to an optimization} of $S_{3}^{2}$ over  all $\rho_{x}^{0}$, $ \hat{b}_{y_1}$ and $ \hat{b}_{y_2}$.
	To optimize $S_{3}^{2} $ we note that the states $\rho_{x}^{0}$ prepared  by Alice should be pure i.e., $||\vec{a}_{x}||=1$. Otherwise the magnitudes $||\vec{m}_{r}||$s decrease leading to an decrement of overall success probability. Again, we must have antipodal pairs (joining vertices of a unit cube inside the Bloch sphere ) constituting each $\vec{m}_{r}$ in Eq. \eqref{eq:effd}. So overall maximization of $S_{3}^{2}$ requires the following;	$\vec{a}_{000}=-\vec{a}_{111}$, $\vec{a}_{001}=-\vec{a}_{110}$, $\vec{a}_{010}=-\vec{a}_{101}$ and $\vec{a}_{100}=-\vec{a}_{001}$. From this choice we can rewrite the $\vec{m}_{r}$s as,  $\vec{m}_1= 2(\vec{a}_{000}+\vec{a}_{001}+\vec{a}_{010}-\vec{a}_{100})$ , $\vec{m}_2= 2(\vec{a}_{000}+\vec{a}_{001}-\vec{a}_{010}+\vec{a}_{100})$ and 	$\vec{m}_3= 2(\vec{a}_{000}-\vec{a}_{001}+\vec{a}_{010}+\vec{a}_{100})$. 
	
	It is seen from Eq. (\ref{eq:succ2}) that as $\alpha_{1}>\beta_{1}$ so for the maximization of $S_{3}^{2}$ we must have $\vec{m}_{r}$ to be along the direction of $ \hat{b}_{y_2}$ when $r =y_2$. We can then write
	
	\begin{equation}
		\begin{split}
			S_{3}^{2} &\leq  \frac{1}{2}+\frac{1}{48}\Big[2\left(\alpha_{1}^2-\beta_{1}^{2}\right)\sum_{r=1,2,3}||\vec{m}_{r}||\\
			&+\frac{4\beta_{1}^2}{3} \sum_{r, y_{1}=1,2,3}\left(\vec{m}_{r}\cdot \hat{b}_{y_{1}}\right)\left( \hat{b}_{y_{1}}\cdot \hat{b}_{r}\right)\Big].  \label{eq:succ211}
		\end{split}
	\end{equation}
	
	{Using concavity of the square root} $\sum_{y_{1}=1}^{3}||\vec{m}_{r}|| \leq \sqrt{3\sum_{y_{1}=1}^{3}||\vec{m}_{r}||^{2}}$ and putting the expressions of $\vec{m}_{r}$s, we have
	\begin{equation}
		\sum_{y_{1}=1}^{3}||\vec{m}_{r}|| \leq 2\sqrt{48-3\left(\vec{a}_{000}-\vec{a}_{001}-\vec{a}_{010}-\vec{a}_{100}\right)^{2}}. \label{eq:msum}
	\end{equation}
	We can then get $max(\sum_{y_{1}=1}^{3}||\vec{m}_{r}||)=8\sqrt{3}$ when the condition 
	\begin{align}
		\label{poco}
		\vec{a}_{000}-\vec{a}_{001}-\vec{a}_{010}-\vec{a}_{100}=0 
	\end{align}
	in Eq. \eqref{eq:msum} is satisfied. This in turn provides each of the $||\vec{m}_{r}||$ is $8/\sqrt{3}$ when equality in Eq. \eqref{eq:msum} holds. The condition in Eq. (\ref{poco}) leads the following relations between the input Bloch vectors, $		\vec{a}_{000}\cdot\vec{a}_{001}+\vec{a}_{000}\cdot\vec{a}_{010}+\vec{a}_{000}\cdot\vec{a}_{100}=1$ , 	\	$\vec{a}_{000}\cdot\vec{a}_{001}-\vec{a}_{001}\cdot\vec{a}_{010}-\vec{a}_{001}\cdot\vec{a}_{100}=1$, \
	$\vec{a}_{000}\cdot\vec{a}_{010}-\vec{a}_{001}\cdot\vec{a}_{010}-\vec{a}_{010}\cdot\vec{a}_{100}=1,$ and  \ $\vec{a}_{000}\cdot\vec{a}_{100}+\vec{a}_{010}\cdot\vec{a}_{100}+\vec{a}_{010}\cdot\vec{a}_{100}=1$.
	Solving the above set of relations one finds $\vec{a}_{000}\cdot\vec{a}_{001}= -\vec{a}_{010}\cdot\vec{a}_{100}$ ,  $\vec{a}_{000}\cdot\vec{a}_{010}= -\vec{a}_{001}\cdot\vec{a}_{100}$ and  $\vec{a}_{000}\cdot\vec{a}_{100}= -\vec{a}_{001}\cdot\vec{a}_{010}$. Further by noting that $||\vec{m}_{1}|| = ||\vec{m}_{2}|| = ||\vec{m}_{3}||$, we get $\vec{a}_{x}\cdot\vec{a}_{x'}=\left(+\frac{1}{3}\right)\delta_{xx'}$ for $x=000, x'\in\{001,010,100\}$ and $\vec{a}_{x}\cdot\vec{a}_{x'}=\left(-\frac{1}{3}\right)\delta_{xx'}$ for other combinations of $x$ and $x'$ where $x'\neq x$ and $\vec{a}_{x}\cdot\vec{a}_{x'}=1$ for $x'= x$. It can be easily checked that the four unit vectors $\vec{a}_{000}$, $\vec{a}_{001},\vec{a}_{010}$ and $\vec{a}_{100}$ form a regular tetrahedron when represented on the Bloch sphere. 
	
	Impinging the above conditions into Eq. \eqref{eq:effd} we immediately get,   $\hat{m}_{r}\cdot \hat{m}_{r^{'}}=\delta_{rr^{'}}$, i.e., $\hat{m}_1$, $\hat{m}_2$ and $\hat{m}_3$ are mutually orthogonal unit vectors. Without loss of generality we can then fix $\hat{m}_1=\hat{x}$, $\hat{m}_2=\hat{y}$ and $\hat{m}_3=\hat{z}$. This implies that the maximum value $max\left(S_{3}^{2}\right)\equiv\Delta_{2}$ can be obtained when the unit vectors of Bob${_2}$ are $ \hat{b}_{y_{2}=1}=\hat{x}$, $ \hat{b}_{y_{2}=2}=\hat{x}$ and $ \hat{b}_{y_{2}=3}=\hat{z}$.  It is then straightforward to understand from the last part of Eq. (\ref{eq:succ211}) that the measurement settings of Bob$_{1}$ has to be same as Bob$_{2}$, $ \hat{b}_{y_{2}}= \hat{b}_{y_{1}}=\hat{m}_{r}$ when $y_{1}=y_{2}$=r. We then have,

	\begin{equation}
		\label{eq:succ2d}
		S_{3}^{2}\leq\frac{1}{2}+\frac{\left(3\alpha_{1}^2-\beta_{1}^2\right)}{6\times24} \sum\limits_{y_{1}=1,2,3}||\vec{m}_{r}||\equiv \Delta_{2}.
	\end{equation}
	It can be seen that given the values of $\alpha_{1}$ and $\beta_{1}$, the maximization of $S_{3}^{2}$ provides the success probability of Bob$_{1}$ {of the form},
	
	\begin{equation}
		S_{3}^{1}=\frac{1}{2}+\frac{\alpha_{1} \beta_{1}}{12}\sum_{y_{1}=1,2,3}||\vec{m}_{r}||\label{eq:succ12}.
	\end{equation}\\

	Note that both $S_{3}^{1}$ and $S_{3}^{2}$ are simultaneously optimized when the quantity $\sum_{y_{1}=1,2,3}||\vec{m}_{r}||$ is optimized. Let us denote $max(S_{3}^{1})$ as $\Delta_{1}$. By putting the values of $\alpha_{1}$ and $\beta_{1}$ and by noting $max(\sum_{y_{1}=1}^{3}||\vec{m}_{r}||))=8\sqrt{3}$, we thus have the optimal pair of success probabilities corresponding to Bob$_{1}$ and Bob$_{2}$ {given by}, 	\begin{eqnarray}
		\Delta_{1} = \frac{1}{2}\left(1+\frac{\lambda_{1}}{\sqrt{3}}\right); \ \
		\Delta_{2}   = \frac{1}{2}\left(1+ \frac{1+2\sqrt{1-\lambda_{1}^2}}{3\sqrt{3}}\right)  \label{eq:succ21}
	\end{eqnarray}

	The optimal trade-off between success probabilities of Bob$_{2}$ and Bob$_{1}$ can then be written as,
	
	\begin{equation}
		{\Delta_{2}({\Delta_{1}})= \frac{1}{2}+\frac{\left(1+2\sqrt{12\Delta_{1}-12\Delta_{1}^{2}-2}\right)}{6\sqrt{3}}},
	\end{equation}
	
	\begin{figure}
		\centering
		\includegraphics[scale=0.47]{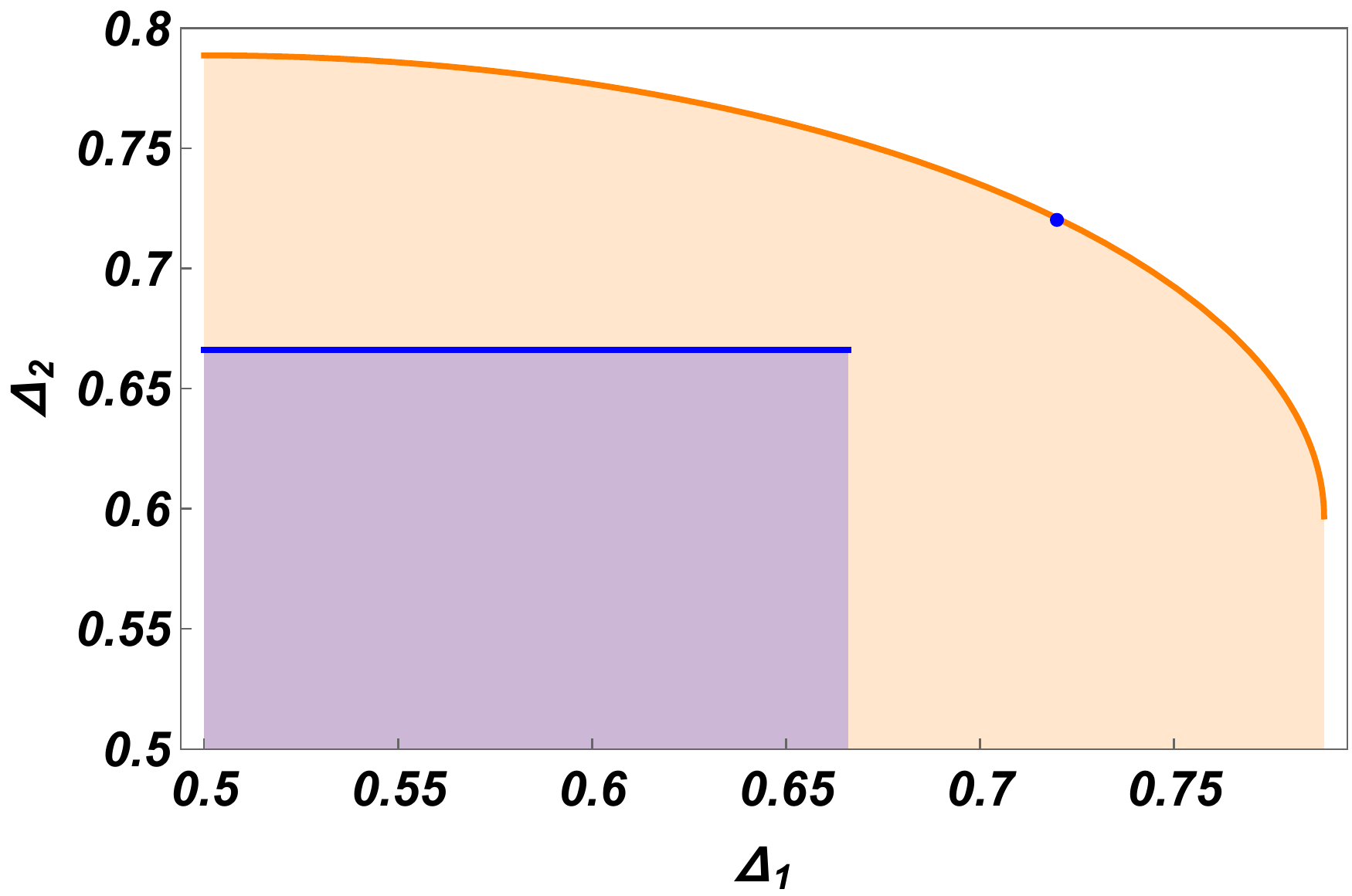}		
		\caption{Optimal trade-off between quantum success probabilities of Bob$_{1}$ and Bob$_{2}$ is shown by solid orange curve while the shaded portion gives the sub-optimal range. Blue solid line is for classical parity-oblivious RAC for the same two observers. } \label{fig3}
	\end{figure} 
	
	yielding that $(\Delta_{2})^{\Delta_{1}}$ is the function of $\Delta_{1}$ only, which is now solely dependent on sharpness parameter $\lambda_{1}$ once the $\Delta_{2}$ is maximized. 
	
	Fig.\ref{fig3} represents the optimal trade-off characteristics between $\Delta_{1}$ and $\Delta_{2}$. The blue line corresponds to the maximum success probability in classical strategy showing no information-disturbance trade-off between the measurements. Each observer can get maximum success probability without any dependence on other observers. On the other hand, the optimal pair $(\Delta_{2}, {\Delta_{1}})$ in quantum theory provide a trade-off. The more the Bob$_{1}$ disturbs the system more the information is gained by him, and $\Delta_{1}$ increases.This eventually decreases the success probability $\Delta_{2}$ of Bob$_{2}$ and vice-versa. The orange curve gives the trade-off for quantum success probabilities, and each point on it certifies a unique value of unsharpness parameter $\lambda_{1}$. For example when Bob$_{1}$ and Bob$_{2}$ gets equal quantum advantage, i.e, $\Delta_{1}=\Delta_{2}=(17+\sqrt{3})/26\approx0.72$, the sharpness parameter of the Bob$_1$ is $\lambda_{1}=(4\sqrt{3}+3)/13 \approx0.763$. This is shown in Fig. \ref{fig3} by the blue point on the orange curve. Similarly, each point in the orange region in Fig.\ref{fig3} certify an unique value of $\lambda_{1}$.

	Thus the optimal pair $(\Delta_{1}, \Delta_{2})$ uniquely certify the preparation of Alice and measurements of Bob$_{1}$ and Bob$_{2}$, and the unsharpness parameter $\lambda_{1}$. The certification statements are the following.
	
	(i) Alice has encoded her message into the eight quantum states that are pure and pairwise antipodal represented by points at the vertices of a unit cube inside the Bloch sphere. One of the examples is, $\vec{a}_{000}=(\hat{x}+\hat{y}+\hat{z})/\sqrt{3}$, $\vec{a}_{001}=(\hat{x}+\hat{y}-\hat{z})/\sqrt{3}$,	$\vec{a}_{010}=(-\hat{x}+\hat{y}+\hat{z})/\sqrt{3}$, $\vec{a}_{100}=(\hat{x}-\hat{y}+\hat{z})/\sqrt{3}$ and their respective antipodal pairs.
	
	(ii) Bob$_{1}$ performs unsharp measurement corresponding to the observables along three mutually unbiased bases, say, $B_{1}=\lambda_{1}\sigma_{x}$, $B_{2}=\lambda_{1}\sigma_{y}$ and $B_{3}=\lambda_{1}\sigma_{z}$ where $\lambda_{1}=\sqrt{3}\left(2\Delta_{1}-1\right)$. Bob$_{2}$'s measurement settings are same as Bob$_{1}$ but sharp measurement of rank one projective value measures.
	
	It is important to note here that the input states those provides the optimal pair ($\Delta_{1},\Delta_{2}$) must satisfy the parity-oblivious condition in quantum theory. Otherwise the comparison between the classical and quantum bounds becomes unfair. For $n=3$ the parity set $\mathbb{P}_{3}$ {contains} four elements $\mathbb{P}_{3}:=\{011,101,110,111\}$. For every element $s\in \mathbb{P}_{3}$ the parity-oblivious condition given by Eq. (\ref{poc}) has to be satisfied. We see that for $s=011$, Alice's input states satisfy the parity-oblivious condition, $\left(\rho_{000}+\rho_{100}+\rho_{011}+\rho_{111}\right)/4=\left(\rho_{010}+\rho_{101}+\rho_{001}+\rho_{110}\right)/4 =\mathbb{I}/2$. This is due to the fact that for the optimal pair ($\Delta_{1}$,$\Delta_{2}$) one requires antipodal pairs, i.e., $\left(\rho_{000}+\rho_{100}\right)/2=\left(\rho_{011}+\rho_{111}\right)/2=\mathbb{I}/2$ and $\left(\rho_{010}+\rho_{001}\right)/2=\left(\rho_{101}+\rho_{110}\right)/2=\mathbb{I}/2$. Similarly, for $s=101$ and $s=110$ the parity-oblivious condition is automatically satisfied and constitute a trivial constraint to Alice's inputs. But for $s=111$, to satisfy parity-oblivious condition Alice's input must satisfy the relation $\left(\rho_{111}+\rho_{100}+\rho_{001}+\rho_{010}\right)/4=\left(\rho_{000}+\rho_{101}+\rho_{011}+\rho_{110}\right)/4 =\mathbb{I}/2$. This demands a non-trivial relation to be satisfied by Alice's inputs is $\vec{a}_{000}-\vec{a}_{001}-\vec{a}_{010}-\vec{a}_{100}=0 $. Interestingly, this is the condition which was required to obtain $\Delta_{2}$ and $\Delta_{1}$ in Eq. (\ref{poco}).  Thus, Alice's input satisfy the parity-oblivious constraint as imposed in the classical RAC. 
	
	\subsection{Sub-optimal scenario}
	Note that there are many practical reasons due to which the precise certification may not be possible in the real experimental scenario. In other words, the optimal pair of success probabilities may not be achieved in a practical scenario, and hence certification of $\lambda_{1}$ may not be possible uniquely. 	We argue that even in sub-optimal scenarios when $S_{3}^{1}\leq \Delta_{1}$ and $S_{3}^{2}\leq \Delta_{2}$ the certification of an interval of values of $\lambda_{1}$ is possible from our protocol.

	From Eq. \eqref{eq:succ21} we find that $\Delta_{1}$ can provide a lower bound $(\lambda_{1})^{min}$ to the unsharpness parameter $\lambda_{1}$ as,
	
	\begin{equation}
		\lambda_{1}\geq \sqrt{3}(2\Delta_{1}-1) \equiv (\lambda_{1})^{min} \label{eq:lowbound3},
	\end{equation}
	and the upper bound $(\lambda_{1})_{max}$ of $\lambda_{1}$ as,
	\begin{equation}
		\lambda_{1} \leq \sqrt{1-\Bigg(\frac{3\sqrt{3}(2\Delta_{2}-1)}{2}-\frac{1}{2}\Bigg)^2}\equiv (\lambda_{1})^{max}\label{eq:upbound3}.
	\end{equation}
	Here we assumed that Bob$_{2}$ performs sharp projecting measurement with $\lambda_{2}=1$. Note that, both the upper and the lower bounds saturate and become equal to each other when $\Delta_{2}$ reaches its maximum value. Now, the quantum advantage for Bob$_1$ requires $\Delta_{1}> 2/3$ which fixes the $(\lambda_{1})^{min}= 1/\sqrt{3}\approx 0.57$ yielding that any value of $\lambda_{1}\in [1/\sqrt{3},1]$ provides quantum advantage for  Bob$_{1}$. 	Again as there is a trade-off between $\Delta_{1}$ and $\Delta_{2}$, in order to obtain the quantum advantage for Bob$_{2}$ the value of $\lambda_{1}$ has the upper bound $(\lambda_{1})^{max}=\sqrt{\sqrt{3}/2}\approx 0.93$.

	Hence, when both Bob$_1$ and Bob$_2$ get quantum advantage the following interval $0.57\leq \lambda_{1}\leq 0.93$ can be certified. {The more the accuracy of the experimental observation the more precise certification of $\lambda_{1}$ is achieved.}   Importantly, this range can be further fine-tuned (in particular the upper bound) if Bob$_3$ gets the quantum advantage. This is explicitly discussed in the Sec. IV and Sec. V. 
	
	\section{Sequential $3$-bit RAC for three or more Bobs}
	Let us now investigate the case when more than two independent Bobs perform the sequential measurement. The question is how many can share quantum advantage sequentially. First, to find the success probability for Bob$_{3}$ we assume that Bob$_{2}$ performs unsharpness measurement with unsharpness parameter $\lambda_{2}$. By using Eq. \eqref{eq:D} again, we calculate the average state received by Bob$_{3}$ is $\rho^{2}_{x}=\left(\mathbb{I}+\vec{a}^{2}_{x}\cdot \sigma\right)/2$, 
	where 
	\begin{eqnarray}
		\vec{a}^{2}_{x}&&= 4\left(\alpha_{1}^2-\beta_{1}^{2}\right)\left(\alpha_{2}^2-\beta_{2}^{2}\right)\vec{a}_{x}+\frac{8}{3}\beta_{1}^{2}\left(\alpha_{2}^2-\beta_{2}^{2}\right) \sum_{y_{1}=1,2,3}\left( \hat{b}_{y_{1}}\cdot\vec{a}_{x}\right) \hat{b}_{y_{1}} \nonumber \\ &&\hspace{1.5cm}+\frac{8}{3}\beta_{2}^{2}\left(\alpha_{1}^2-\beta_{1}^{2}\right) \sum_{y_{2}=1,2,3}\left( \hat{b}_{y_{2}}\cdot\vec{a}_{x}\right) \hat{b}_{y_{2}} \\
		\nonumber
		&&\hspace{1.5cm}+\frac{16\beta_{1}^2\beta_{2}^{2}}{9}\sum_{y_{1},y_{1}=1,2,3}\left( \hat{b}_{y_{1}}\cdot\vec{a}_{x}\right)\left( \hat{b}_{y_{1}}\cdot \hat{b}_{y_{2}}\right) \hat{b}_{y_{2}}.
	\end{eqnarray}
	
	The quantum success probability for Bob$_{3}$ can then be written as,

	\begin{eqnarray}
		S_{3}^{3} =&&\frac{1}{2}+ \frac{1}{48} Tr\Bigg[ 4\left(\alpha_{1}^2-\beta_{1}^{2}\right)\left(\alpha_{2}^2-\beta_{2}^{2}\right)\sum_{y_{3}=1,2,3}\left(\delta_{r,y_{3}}\vec{m}_{r}\cdot \hat{b}_{y_{3}}\right) \nonumber \\ 
		&&+\frac{8}{3}\beta_{1}^{2}\left(\alpha_{2}^2-\beta_{2}^{2}\right) \sum_{y_{1},y_{3}=1,2,3}\left( \delta_{r,y_{1}}\vec{m}_{r}\cdot \hat{b}_{y_{1}}\right)\left( \hat{b}_{y_{1}}\cdot \hat{b}_{y_{3}}\right) \nonumber \\ 
		&&+\frac{8}{3}\beta_{2}^{2}\left(\alpha_{1}^2-\beta_{1}^{2}\right) \sum_{y_{2},y_{3}=1,2,3}\left( \delta_{r,y_{2}}\vec{m}_{r}\cdot \hat{b}_{y_{2}}\right)\left( \hat{b}_{y_{2}}\cdot \hat{b}_{y_{3}}\right) \nonumber\\
		&&+\frac{16\beta_{1}^2\beta_{2}^{2}}{9}\sum_{y_{1},y_{2},y_{3}=1,2,3}\left( \delta_{r,y_{1}}\vec{m}_{r}\cdot \hat{b}_{y_{1}}\right)\left( \hat{b}_{y_{1}}\cdot \hat{b}_{y_{2}}\right) \left( \hat{b}_{y_{2}}\cdot \hat{b}_{y_{3}}\right) \Bigg]. \label{eq:succ3}
	\end{eqnarray}
	
	
	Using the maximization scheme presented in Sec. III, it is straightforward to obtain that $m_{y_{1}}$ must be in the same direction of $\hat{b}_{y_{3}}$ when $r$=$y_{3}$ and $\hat{b}_{y_{1}}\cdot\hat{b}_{y_{2}}$ when $y_{1}=y_{2}$ and so on. We can then write,
	\begin{equation}
		S_{3}^{3} \leq \frac{1}{2}+\frac{1}{6\times72}\Big[\left(3\alpha_{1}^2-\beta_{1}^2\right)\left(3\alpha_{2}^2-\beta_{2}^2\right)\sum_{y_{1}=1,2,3}||\vec{m}_{r}||\Big]\equiv \Delta_{3},\label{eq:succ3d}
	\end{equation}
	where the maximum value $max(S_{3}^{3})= \Delta_{3}$. It can then be seen that $S_{3}^{3}$, $S_{3}^{2}$ and $S_{3}^{1}$ can be jointly optimized if the quantity $\sum_{y_{1}=1}^{3}||\vec{m}_{r}||$ is maximized. 	By using the fact $\max({\sum_{y_{1}=1}^{3}||\vec{m}_{r}}||)=8\sqrt{3}$  and putting the values of $\alpha_{1}, \alpha_{2}, \beta_{1}$ and $\beta_{2}$ we have
	
	\begin{equation}
		\begin{split}
			\Delta_{3}  =\frac{1}{2}\Bigg(1+ \frac{(1+2\sqrt{1-\lambda_{1}^2})(1+2\sqrt{1-\lambda_{2}^2})}{9\sqrt{3}}\Bigg). \label{eq:succ31}
		\end{split}
	\end{equation}

	Using the values of $\Delta_{1}$ and $\Delta_{2}$ from Eq. \eqref{eq:succ21}, by combining them in Eq. \eqref{eq:succ31} and further simplifying we find the optimal triple between success probabilities of Bob$_{1}$, Bob$_{2}$ and Bob$_{3}$,  is given by,
	
	\begin{equation}
		\Delta_{3} ^{\Delta_{1},\Delta_{2}}=\frac{1}{2}\Bigg\{1+\frac{\Xi_{1}+2\sqrt{\Xi_{1}^{2}-\Xi_{2}^{2}}}{9\sqrt{3}}\Bigg\}, \label{eq:opt}
	\end{equation}
	where $\Xi_{1}=1+2\sqrt{12\Delta_{1}-12\Delta_{1}^2-2}$ and  $\Xi_{2}=3\sqrt{3}(2\Delta_{2}-1)$.\\
	
	This becomes nontrivial when at least one of the $\Delta_{1}$, $\Delta_{2}$ or $\Delta_{3}$ surpasses classical bound, as discussed earlier. 
	

	The above approach can be generalized for $k^{th}$ number of Bobs where $k$ is arbitrary. When Bob$_{k}$ uses optimal choice of observables, the average state relayed to $(k+1)^{th}$ Bob can be written as $\rho_{x}^{k}=\frac{1}{2}\Bigg(\mathbb{I}+\vec{a}_{x}^{k}\cdot \sigma)\Bigg)$, where,
	\begin{equation}
		\vec{a}_{x}^{k}=\prod_{i=1}^{k}\frac{(1+2\sqrt{1-\lambda_{i}^2})\vec{a}_{x}^{0}}{3^{k}}.
	\end{equation}
	Here $ \vec{a}_{x}^{0}$s are the Bloch {vectors}  of the states prepared by Alice. Simply, this state is along the same direction of the initial state prepared by Alice, but the length of the the Bloch vector is shrunk by an amount, $\prod_{i=1}^{k}(1+2\sqrt{1-\lambda_{i}^2})/3^{k}$. The optimal success probability of $k^{th}$ Bob can be written as 
	\begin{equation}
		\Delta_{k}=\frac{1}{2}\Bigg[1+{\frac{1}{3^{k}} \prod_{i=1}^{k-1}\sqrt{3}(1+2\sqrt{1-\lambda_{i}^2})}\Bigg], \label{eq:anyk}
	\end{equation}

	where k$^{th}$ Bob performs sharp projective measurement. In order to examine this the longest sequence to which all Bobs can get quantum advantage, let us  consider the situation when Bob$_{1}$, Bob$_{2}$ and Bob$_{3}$ implement their unsharp measurements with lower critical values of unsharpness parameters at their respective sites. From Eq. \eqref{eq:anyk}, we get the lower critical values of $\lambda_{1}=\frac{1}{\sqrt{3}}=0.5773$, $\lambda_{2}=0.6578$ and $\lambda_{3}=0.7873$. Putting these values in Eq. \eqref{eq:anyk}  we get $\Delta_{4}= 0.6575 < 2/3$ for $k=4$, i.e., no quantum advantage is obtained for the fourth Bob.

	\section{Certification of multiple independent unsharpness parameters}
	
\begin{figure}
		\centering
		\includegraphics[scale=0.3]{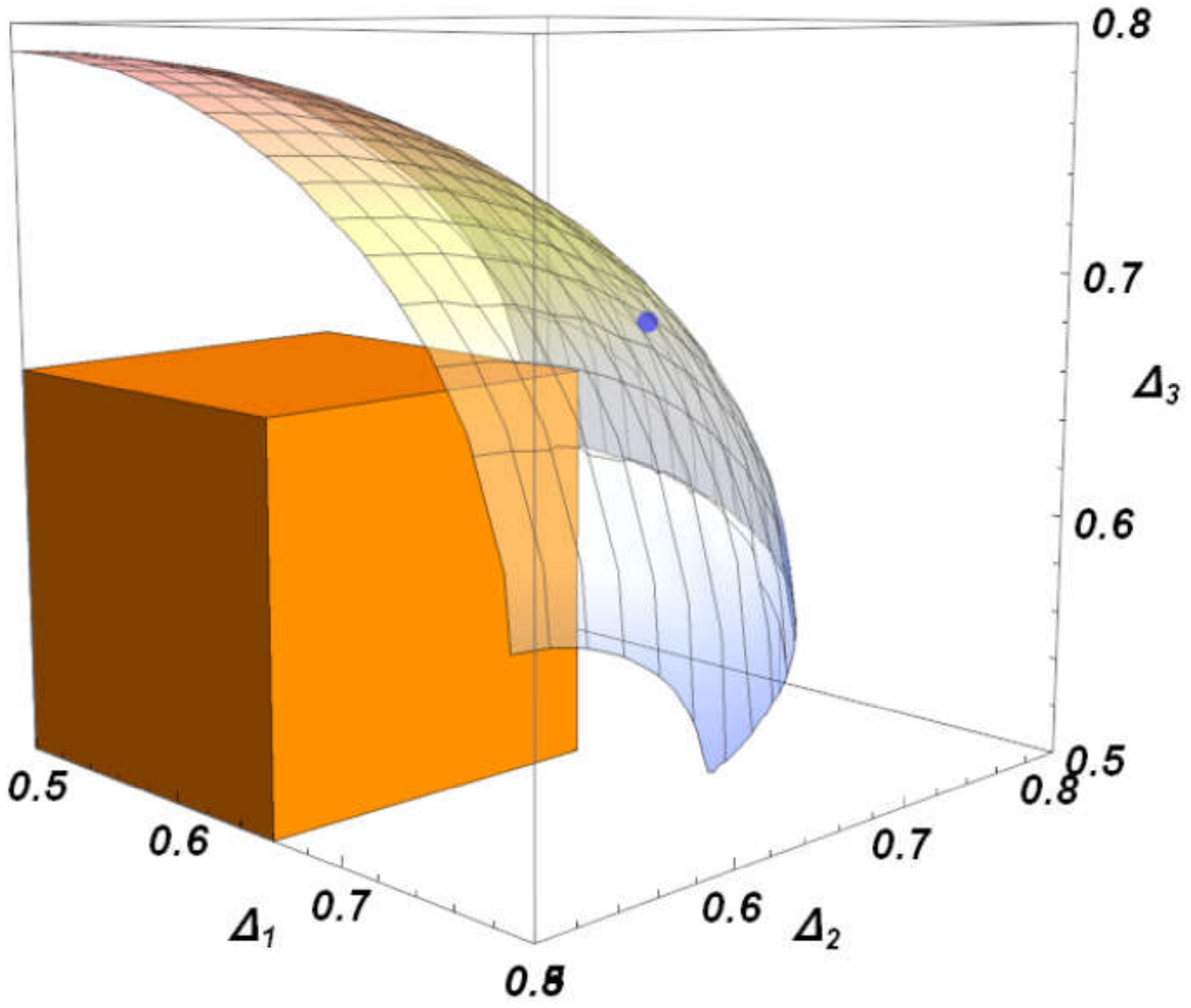}
		\caption{Optimal trade-off between success probabilities for Bob$_{1}$, Bob$_{2}$ and Bob$_{3}$. Blue point on the three dimensional graph indicates the point where the success probabilities coincide, i.e,  $\Delta_{1}$ = $\Delta_{2}$ =  $\Delta_{3}$ = 0.686 $\geq$ $\frac{2}{3}$.}\label{fig2}
	\end{figure}

	Optimal triple of the success probabilities of Bob$_{1}$, Bob$_{2}$ and Bob$_{3}$ are given by Eq. \eqref{eq:opt} is ploted in Fig.\ref{fig2}.	The orange cube represents the classical success probabilities showing no trade-off. The three dimensional semi-paraboloid  over the cube represents quantum trade-off between success probabilities. Each point on the surface of semi-paraboloid in Fig.\ref{fig2} uniquely certify $\lambda_{1}$ and $\lambda_{2}$ while we take measurement of Bob$_{3}$ is projective having $\lambda_{3}=1$.  For instance, let us consider the point when all the success probabilities have the equal value, $(\Delta_{1}, \Delta_{2}, \Delta_{3})=(0.686,0.686,0.686)$. This particular point uniquely certify $\lambda_{1}=0.6443$ and $\lambda_{2}=0.7641$. We can take any arbitrary point, for example, $(\Delta_{1}, \Delta_{2}, \Delta_{3})=(0.67,0.675,0.704)$ on the graph. This point certifies the unsharpness parameters $\lambda_{1}=0.588$ and $\lambda_{2}=0.695$. 
	
	In principle, one can uniquely certify unsharpness parameters $\lambda_{1}$ and $\lambda_{2}$ through the optimal triple but in practical scenario quanum instruments are subjected to imperfections and losses. We show that our scheme can also certify a range of values of  $\lambda_{1}$ and  $\lambda_{2}$. When all the three Bobs get quantum  advantage, the optimal values of success probabilities fix the range of $\lambda_{1}$ and  $\lambda_{2}$ obtained from Eqs. \eqref{eq:succ21} and \eqref{eq:succ31} are respectively given by
	\begin{equation}
		\sqrt{3}(2\Delta_{1}-1) \leq \lambda_{1} \leq \sqrt{1-\Bigg(\frac{3\sqrt{3}(2\Delta_{2}-1)}{2\lambda_{2}}-\frac{1}{2}\Bigg)^2}\label{eq:bound1}
	\end{equation}
	and
	\begin{equation}
		\frac{3\sqrt{3}(2\Delta_{2}-1) }{1+2\sqrt{1-\lambda_{1}^2}}\leq \lambda_{2} \leq \sqrt{1-\frac{1}{4}\Bigg(\frac{9\sqrt{3}(2\Delta_{3}-1)}{1+2\sqrt{1-\lambda_{1}^2}}-1\Bigg)^2}. \label{eq:bound2}
	\end{equation}
	It is seen from Eq. \eqref{eq:bound1} that although the lower bound of $\lambda_{1}$ depends only on the observational statistics $\Delta_{1}$, the upper bound does not. Rather, the upper bound of $\lambda_{1}$ is a function of $\lambda_{2}$ and the optimal success probability $\Delta_{2}$ of Bob$_{2}$. On the other hand, Eq. \eqref{eq:bound2} shows that both the upper and lower bound of $\lambda_{2}$ are not only dependent on the observational statistics but also on $\lambda_{1}$. This interdependence of sharpness parameters suggests a trade-off between them. By taking a value of $\lambda_{2}$, which is just above the lower critical value, we sustain the quantum advantage to subsequent Bobs. Each value of $\lambda_{1}$ above the lower critical value fixes the minimum value of unsharpness that is required to get the quantum advantage for Bob$_{2}$ while the need for a sustainable advantage to a subsequent Bob fixes its upper bound. The whole program runs as follows.

	As already mentioned, the minimum value of unsharpness parameter required to get a quantum advantage is $\lambda_{min}=\frac{1}{\sqrt{3}}$. So a value $\lambda_{1}= \frac{1}{\sqrt{3}}+\epsilon_{1}$ where $\epsilon_{1}> 0$ would suffice. Furthermore putting this value of $\lambda_{1}$ into Eq. \eqref{eq:succ21} we can estimate the minimum value  of unsharpness in Bob$_{2}$'s measurement as $\lambda_{2}= 0.6578+ \epsilon_{2}$ for all $\epsilon_{2} >0.3533\epsilon_{1} + O(\epsilon_{1}^2)$. Similarly, to get advantage for Bob$_{3}$ the minimum value of unsharpness parameter can be estimated from Eq. \eqref{eq:succ31} as $\lambda_{3}=0.7873+\epsilon_{3}$ where $\epsilon_{3}=0.1187\epsilon_{1}+O(\epsilon_{1}^2)$.

	For an explicit example of how to estimate the interval of values of unsharpness parameters, let us consider the task where Bob$_{1}$ chooses the optimal measurement settings with unsharpness $\lambda_{1}=\frac{1}{\sqrt{3}}+0.05=0.6273$.
	Now their task is to set the unsharpness parameter such that Bob$_{2}$ and Bob$_{3}$ can get the quantum advantage. From Eq. \eqref{eq:bound2} Bob$_{2}$ can set the unsharpness parameter any value in the range $0.6772\leq\lambda_{2}\leq0.8566$. The minimum value of unsharpness parameter required to get an advantage at Bob$_{3}$'s site depends on both $\lambda_{1}$ and $\lambda_{2}$. With the same $\lambda_{1}$, for a lower critical value of $\lambda_{2} (=0.6772)$, the unsharpness parameter of Bob$_{3}$'s measurement is lower bounded as $0.8220\leq\lambda_{3}$, on the other hand when Bob$_{2}$ performed his measurement with upper critical unsharpness ($\lambda_{2}=0.8566$) the bounds on $\lambda_{3}$ becomes unity, which comes directly from Eq. \eqref{eq:succ31}.

	Due to the fact that when Bob$_{3}$ gets the quantum advantage, the upper bound of $\lambda_{1}$ depends on both $\lambda_{2}$ and the success probabilities of respective Bobs,  $(\lambda_{1})^{max}$ is more restricted than the one obtained when quantum advantages only for Bob$_{1}$ and Bob$_{2}$ were considered.  Using Eq. \eqref{eq:bound1} by putting $\Delta_{2}=2/3$ and $\lambda_{2}=1$ we had earlier fixed the allowed interval of $\lambda_{1}$ as, $0.57 \leq \lambda_{1} \leq 0.93$  when only two Bobs are able to get advantage. This is discussed in Sec. III.  Now if Bob$_{3}$ gets quantum advantage ( $\Delta_{3}>2/3$) along with Bob$_{1}$ and Bob$_{2}$, then using Eq. \eqref{eq:succ21} and Eq. \eqref{eq:succ31}  we get the the interval of $\lambda_{1}$ has to be $0.57 \leq \lambda_{1} \leq 0.77$. Thus, the quantum advantage extended to the Bob$_{3}$ provides a narrower interval $\lambda_{1}$ by decreasing the upper bound. In such case more efficient experimental verification is required to test it.

	Let us denote these two ranges by \emph{R$_{1}$} and \emph{R$_{2}$} respectively. To get a practical impression of how such a certification works, let us consider a {seller} who sells a measurement instrument and claims that it works with a particular noise given by its unsharpness parameter $\lambda_{1}$ that belongs to a value within \emph{R$_{2}$}. We can check by a sequential arrangement discussed above and by observing the statistics of Bob$_{3}$ whether the instrument is trusted or not. If we see that Bob$_{3}$ is getting a quantum advantage, then we conclude that the seller is trusted and $\lambda_{1}$ must lie in \emph{R$_{2}$}. On the other hand, a failure of Bob$_{3}$ in getting an advantage will compel us to decide that the instrument is not trusted and $\lambda_{1}$ lie in \emph{R$_{1}$} which may or may not be useful for a particular purpose.
	
	It is then natural to think that if more numbers of Bobs get the quantum advantage, then the interval of $\lambda_{1}$ will be narrower. However, for 3-bit RAC, a fourth observer cannot get the quantum advantage. In search of such a possibility, we consider the $4$-bit sequential RAC. 
	
	\section{4-bit sequential RAC }
	
	In $4$-bit sequential quantum RAC Alice now has a length-4 string randomly sampled from $x\in \{0,1\}^{4}$ which she encodes into sixteen qubits states  $\rho_x = \frac{\mathbb{I}+\vec{a}_{x}\cdot\sigma}{2}$, and {sends} them to Bob$_{1}$. After receiving the system, Bob$_{1}$ randomly performs the measurements of dichotomic observables $B_{y_{1}}$ with $y_{1}\in[4]$ and relay the system to Bob$_{2}$ and so on. This process goes on as long as $k^{th}$ Bob gets the quantum advantage, as also discussed for the 3-bit case. The main purpose of extending our analysis to the $4$-bit case is to examine whether more than three Bobs can get quantum advantage sequentially. This may then provide the certification of more than two unsharpness parameters and more efficient fine-tuning of the interval of values of unsharpness parameter of a given measurement instrument. 
	
	For $4$-bit sequential quantum RAC, we found that for the qubit system, the optimal quantum value of success probability is different in standard and parity-oblivious RAC. Note here that the classical success probability is also different in standard and parity-oblivious RAC. For the latter case, the quantum advantage can be obtained for two Bobs, but for the former case, the quantum advantage cannot exceed one Bob. However, we demonstrate that instead of a single qubit system, if the input states are encoded in a two-qubit system, then four Bobs can get the quantum advantage in parity-oblivious RAC. In fact, in that case, the input two-qubit states required for achieving optimal quantum value naturally satisfy the parity-oblivious conditions for the $4$-bit case. This feature is similar to the $3$-bit case.

	\subsection{Standard $4$-bit RAC with qubit input states}
	We first consider the case when the parity-oblivious constraint is not imposed, and input states are qubit. In that case the  success probability for $4$-bit classical RAC is bounded by $S_{4}^{k}\leq \frac{11}{16}$ \cite{pan2020}. 
	In quantum RAC, we first consider the input states are in the qubit system. We have taken slightly different calculation steps than that of the $3$-bit case (Sec. II), as presented in the Appendix. The quantum success probability for  Bob$_{1}$ can be written as 
	\begin{equation}
		\begin{split}
			S_{4}^{1} &=\frac{1}{64}\sum_{x \in \{0,1\}^4}\sum_{y_{1}=1}^{4}Tr\Big[\rho_{x}^{0} B_{x_{y_{1}}|y_{1}}\Big]\\ 
			&=\frac{1}{2}+\frac{\alpha_{1}\beta_{1}}{16 }\sum_{y_{1}=1}^{4} (-1)^{x_{y_{1}}^{i}}\vec{a}_{x^{i}}\cdot \hat{b}_{y_{1}}. \label{eq:Succ41m}
		\end{split}
	\end{equation}
	Here $x_{y_{1}}^{i}$ is the $y_{1}^{th}$ bit of the $4$-bit input string $x^{i}\in \{0,1\}^{4}$  with first bit $0$, i.e., $x_{1}^{i}=0$ for all $i\in [8]$.   
	Let us now analyze the case when Bob$_{2}$ can get quantum advantage. We find the maximum success probability for Bob$_{2}$ from \eqref{eq:Succ42b} is achieved when $\hat{b}_{1}\cdot\hat{b}_{2}=\hat{b}_{1}\cdot\hat{b}_{3}=\hat{b}_{2}\cdot\hat{b}_{3}=0$ and $\hat{b}_{3}\cdot\hat{b}_{4}=1$. As presented in detail in the Appendix the maximum success probability for the Bob$_{2}$ $max(S_{4}^{2})=\Omega_{2}$	 can be written as 
	\begin{equation}
		max(S_{4}^{2}) = \Omega_{2} =\frac{1}{2}+ \lambda_{2}\Bigg( \frac{(\sqrt{2}+\sqrt{6})}{64}\Bigg)\big(1+3\sqrt{1-\lambda_{1}^2}\big). \label{eq:Succ421m}
	\end{equation}
	\begin{figure}
		\centering
		\includegraphics[scale=0.47]{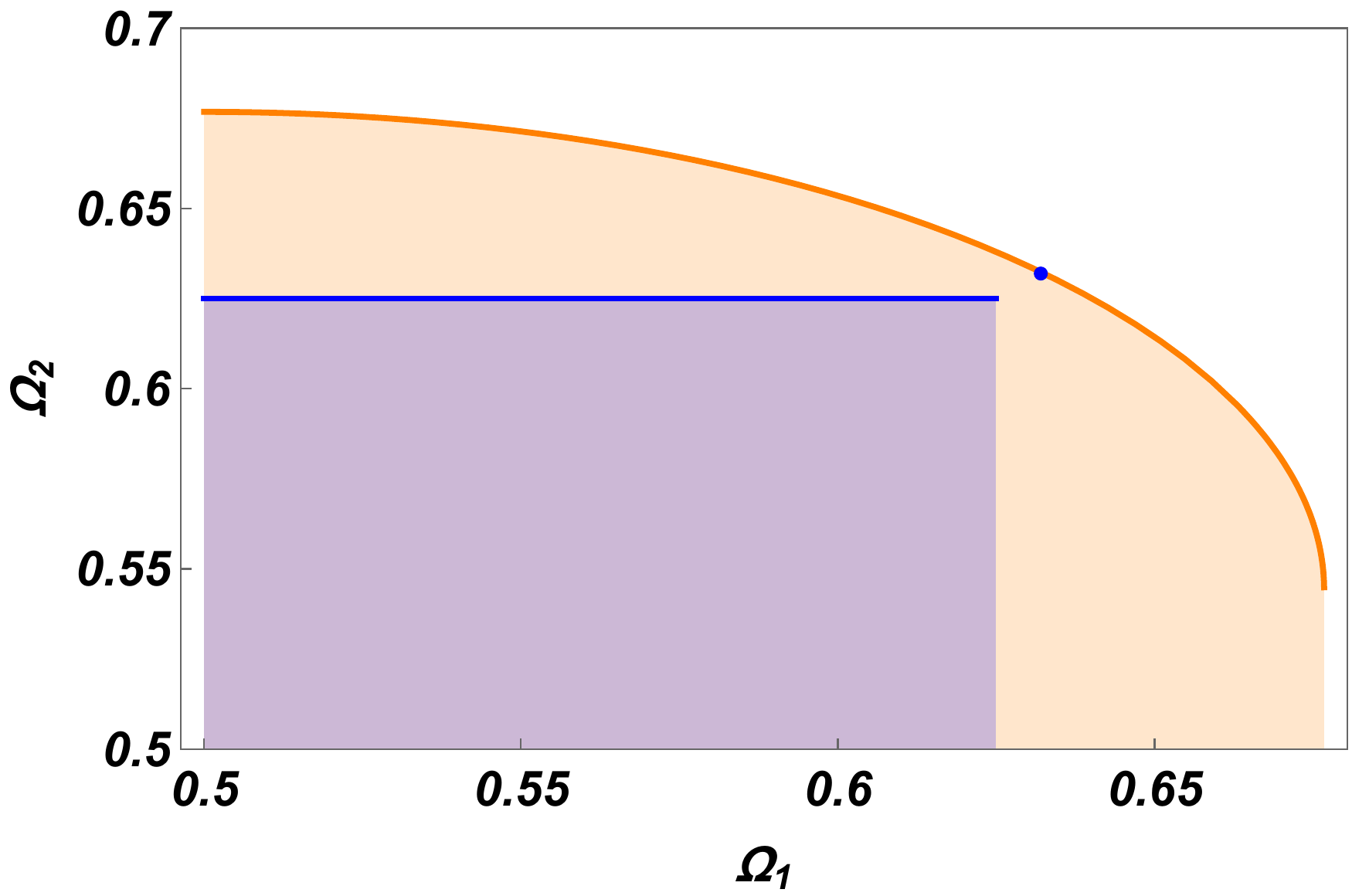}		
		\caption{Optimal trade-off between quantum success probabilities of Bob$_{1}$ and Bob$_{2}$ for 4-bit sequential RAC is shown by solid orange curve while the shaded portion gives the sub-optimal range. Blue solid line is classical success probability for the same two observer constrained by parity-oblivious condition.}	\label{fig4}
	\end{figure} 
	The directions of the Bloch vectors $a_{x}$ and $ \hat{b}_{y_{1}}$ and relations between the different Bloch vectors $\vec{a}_{x}$ are explicitly provided in the Appendix. It is seen that conditions for maximization of $S_{4}^{2}$ also leads the maximum value $\Omega_{1}=max(S_{4}^{1})$ of Bob$_{1}$ in  Eq. \eqref{eq:Succ41m}, which can be written as,
	
	\begin{equation}
		max(S_{4}^{1}) = \Omega_{1} =\frac{1}{2}+ \lambda_{1}\Bigg( \frac{(\sqrt{2}+\sqrt{6})}{16}\Bigg).\label{eq:Succ411m}
	\end{equation}\\
	
	From Eqs. \eqref{eq:Succ41m} and \eqref{eq:Succ421m} we found that to get simultaneous quantum advantage for Bob$_{1}$ and Bob$_{2}$ the minimum values of the unsharpness parameters are $\lambda_{1}^{min}=0.776$ and $\lambda_{2}^{min}=1.074$ respectively. Since the allowed value of $\lambda_{2}$ is $0\leq \lambda_{2}\leq 1$, it is evident that Bob$_{2}$ cannot get quantum advantage. 
	
	\subsection{$4$-bit RAC with qubit input states with parity-oblivious constraint}
	
	We now show that when the parity-oblivious constraint is imposed on Alice's encoding, two sequential Bobs can get quantum advantage {which} enables certification of the unsharpness parameter of the Bob$_1$'s instrument even when the input states are qubit. In such case, the classical preparation non-contextual bound on success probabilities is $(S_{4}^{k})\leq\frac{5}{8}\approx 0.625$. As outlined in Appendix A, respective maximum success probabilities for Bob$_{1}$ and Bob$_{2}$ are

	\begin{equation}
		max(S_{4}^{1}) = \Omega_{1} =\frac{1}{2}+ \lambda_{1}\Bigg( \frac{1}{4\sqrt{2}}\Bigg)\label{eq:Succ41m1}
	\end{equation}
	
	and 	
	\begin{equation}
		max(S_{4}^{2}) = \Omega_{2} =\frac{1}{2}+ \lambda_{2}\Bigg( \frac{\big(1+3\sqrt{1-\lambda_{1}^2}\big)}{16\sqrt{2}}\Bigg). \label{eq:Succ42m1}
	\end{equation}

	The optimal trade-off between the success probabilities of Bob$_{1}$ and Bob$_{2}$ (by considering $\lambda_{2}=1$), is given by
	\begin{equation}
		{\Omega_{2}({\Omega_{1}}) =\frac{1}{2}+\frac{\left(1+3\sqrt{1-8(2\Omega_{1}-1)^{2}}\right)}{16\sqrt{2}}}
		\label{tf4}
	\end{equation}
	
	In Fig.\ref{fig4} we plotted $\Omega_{2}$ against $\Omega_{1}$ which gives the optimal trade-off relation between successive success probabilities. The solid orange curve provides an optimal trade-off, while the shaded portion gives the sub-optimal range. The solid blue line is for the classical case for the same two observers, and there is no trade-off.  As explained for the $3$-bit case, one can uniquely certify the unsharpness parameter from this trade-off relation. For example, when $\Omega_{2}={\Omega_{1}}=0.6805$ the unsharpness parameter is {$\lambda_{1}=\frac{4+6\sqrt{6}}{25}\approx0.7478$. This point is shown in Fig.\ref{fig4} by the black dot on the solid orange curve.
		
		Thus the optimal pair $(\Omega_{1}, \Omega_{2})$ self-test the preparation of Alice, the measurement settings of Bob$_1$ and Bob$_2$, and the unsharpness parameter $\lambda_{1}=2\sqrt{2}\left(2\Omega_{1}-1\right)$. The self-testing statements similar to the $3$-bit case can also be made here.
		
		In sub-optimal scenario, by taking $\lambda_{2}=1$, the Bob$_{1}$ and Bob$_{2}$ both get quantum advantage for a interval of values of $\lambda_{1}$ is given by,
		
		\begin{equation}
			2\sqrt{2}(2\Omega_{1}-1) \leq \lambda_{1} \leq \sqrt{1-\Bigg(\frac{8\sqrt{2}(2\Omega_{2}-1)}{3}-\frac{1}{3}\Bigg)^2}\label{eq:bound41}
		\end{equation}
		
		This above range is calculated by using Eqs.\eqref{eq:Succ41m1} and \eqref{eq:Succ42m1}. Putting the classical bounds of $\Omega_{1}=\Omega_{2}=5/8$, the numerical values of the interval of $\lambda_{1}$ is calculated as $	0.707 \leq \lambda_{1} \leq 0.792$. The more accurate the experimental observation, the more precise the $\lambda_{1}$ can be certified. This implies that an efficient experimental verification is required to confine the $\lambda_{1}$ in a narrow region. Note that the above interval can be made narrower if Bob$_{3}$ gets the quantum advantage. But we found that not more than two Bobs can get the advantage in this case, as given in the Appendix.
		
		\subsection{$4$-bit RAC with two-qubit input states}	
		
		We provide a sketch of the argument here. If in 4-bit parity oblivious RAC the input states are a two-qubit system, then four sequential Bobs can share the quantum advantage (see Appendix), and thus three unsharpness parameters can be certified. Note that the input states providing the optimal quantum value of the success probabilities satisfy the parity-oblivious constraints. The optimal trade-off relation between Bob$_{1}$ and Bob$_{2}$ can be derived by using the Eqs. 	\eqref{eq:Succtwo1}-\eqref{eq:Succtwok}.  
		
		In the sub-optimal scenario, the interval within which the unsharpness parameter is confined becomes much narrower. We argue again that this requires sophisticated experimental realization. When only Bob$_{1}$ and Bob$_{2}$ get quantum advantage, the interval of $\lambda_{1}$ in sub-optimal  scenario is $	0.517 \leq \lambda_{1} \leq 0.934$ and when Bob$_{1}$, Bob$_{2}$ and Bob$_{3}$ get the quantum advantage, the interval becomes $	0.517 \leq \lambda_{1} \leq 0.807 $. However, four Bobs get quantum advantage, the interval of $\lambda_{1}$becomes further narrower $0.517 \leq \lambda_{1} \leq 0.663 $. This then shows that the range of $\lambda_{1}$ becomes more narrow when four sequential Bobs share quantum advantage compared to the case when only the first two Bobs get the quantum advantage.  Note that the analytic expressions can be straightforwardly calculated using Eq.\eqref{eq:Succtwok}. To avoid clumsiness, we skipped them and provided only numerical values.

		\section{Summary and Discussion}
		
		In summary, based on a communication {game} known as the parity-oblivious RAC, we provided two SDI protocols in the prepare-measure scenario to certify multiple unsharpness parameters. Such a certification is demonstrated by examining the quantum advantage shared by multiple sequential observers (we denote Bob$_{k}$ for $k^{th}$ observer). First, we demonstrated that for $3$-bit parity-oblivious sequential RAC, a maximum of three sequential Bobs gets the quantum advantage over the classical preparation non-contextual strategies. We showed that the optimal pair of quantum success probabilities between Bob$_{1}$ and Bob$_{2}$ self-test the qubit states and the measurements of Bob$_{1}$ and  {Bob$_{2}$}. Our treatment clearly shows that in the $3$-bit case, the prepared states providing the optimal quantum success probability satisfy the parity-oblivious restriction. The optimal triple of success probabilities self-test the unsharpness parameters $\lambda_{1}$ and $\lambda_{2}$ of Bob$_{1}$ and Bob$_{2}$ respectively. In realistic experimental scenarios, loss and imperfection are inevitable, and the success probabilities become sub-optimal. In such a case, when only Bob$_{1}$ and Bob$_{2}$ get the quantum advantage, we derived a certified interval of $\lambda_{1}$ within which it has to be confined. The more the precision of the experiment, the more accurate the bound to $\lambda_{1}$ can be certified. We showed that when the quantum advantage is extended to Bob$_{3}$, the interval of $\lambda_{1}$ will be narrower, thereby requiring more efficient experiment realization. 
		
		We extended our treatment to the $4$-bit case. We found that the qubit system cannot achieve the global optimal quantum value of success probability. But if the prepared states are two-qubit systems, we obtain the optimal quantum success probability. The qubit input states for which the maximum success probability is attained do not naturally satisfy the parity-oblivious constraints on input states. We studied three different cases in $4$-bit sequential RAC. First, when input states are qubit, and parity-oblivious constraints are not imposed. In such a case, we have standard RAC, and we found the quantum advantage cannot be extended to Bob$_2$. Thus, this result can be used to certify the states and measurements but cannot certify the unsharpness parameter. We then considered the case when input states are qubit and imposed the parity-oblivious constraints. We showed that, in this case, Bob$_{1}$ and Bob$_{2}$ can share the quantum advantage. We derived a trade-off relation between the success probabilities between Bob$_{1}$ and Bob$_{2}$. Consequently, we demonstrated the certification of the unsharpness parameter of Bob$_{1}$.  In the third case, when the input states are in a two-qubit system, the input states maximize the quantum success probability naturally satisfy the parity-oblivious conditions. In this case, the quantum advantage is extended to four sequential Bobs, and hence three unsharpness parameters can be certified. However, we provided a sketch of this argument, and details of it will be published elsewhere.

		The MTB \cite{mohan2019} protocol of $2$-bit sequential quantum RAC  for certifying the unsharpness parameter has recently been experimentally tested in \cite{anwar2020,fole2020,xiao}. Our protocols can also be tested using the existing technology already adopted in \cite{anwar2020,fole2020,xiao}. Finally, our protocol can also be extended to arbitrary-bit sequential RAC or other parity-oblivious communication games \cite{pan21} which may demonstrate the quantum advantage to more independent observers. This will eventually pave the path for certifying more number of unsharpness parameters. In such a case, in a sub-optimal scenario, the interval of unsharpness parameter of the first Bob will be very narrow. Efficient experimental realization is required to certify such an interval. Study along this line could be an exciting avenue for future research. We also note here that in the other interesting work \cite{miklin2020} along this line, the authors have used the $2$-bit sequential RAC as a tool for certifying the unsharpness parameter. However, the approach adopted in \cite {miklin2020} is different from the MTB protocol. It would then be interesting to extend their certification protocol for $n$-bit RAC or other parity-oblivious communication games. This also calls for further study.
		
		\section{ACKNOWLEDGMENTS}
		The authors acknowledge the support from the project DST/ICPS/QuST/Theme-1/2019/4.
		
		\begin{widetext}
			\appendix
			\section{Details of the calculation for $4$-bit sequential quantum RAC}
			For the $4$-bit case, we provide three conditional maximizations. i) The case when input states are qubit, and parity-oblivious constraints are not imposed. ii) The case when input states are qubit, and parity-oblivious constraints on the input state are imposed. iii) When the input states are a two-qubit system. In this case, the input states maximize the success probability naturally satisfy the parity-oblivious conditions.  
			\subsection{Maximum success probability for qubit system without imposing parity-oblivious constraints }
			In the 4-bit scenario as Alice has encoded her 4-bit string into sixteen quantum states, and Bob$_{1}$ performs four random measurements. From Eq.\eqref{succ1} of the main text, the quantum success probability of Bob$_{1}$'s site is given by:
			\begin{eqnarray}
				S_{4}^{1} =\frac{1}{64}\sum_{x\in \{0,1\}^{4}}\sum_{y_{1}=1}^{4}Tr\Big[\rho_{x} B_{x_{y_{1}}|y_{1}}\Big]=\frac{1}{2}+\frac{\alpha_{1}\beta_{1}}{32}\sum_{x\in \{0,1\}^{4}}\sum_{y_{1}=1}^{4} (-1)^{x_{y_1}}\vec{a}_{x}\cdot \hat{b}_{y_{1}}\big), \label{eq:Succ41}
			\end{eqnarray}
			
			which is Eq. \eqref{eq:Succ41m} in the main text. We denote $m_{y_{1}}=(-1)^{x_{y_1}}\vec{a}_{x}$ which are unnormalized vectors and can explicitly be written as  \\
			
			$ \vec{m}_1=(\vec{a}_{0000}-\vec{a}_{1111})+(\vec{a}_{0001}-\vec{a}_{1110})+(\vec{a}_{0010}-\vec{a}_{1101})+(\vec{a}_{0100}-\vec{a}_{1011})+(\vec{a}_{0111}-\vec{a}_{1000})+(\vec{a}_{0011}-\vec{a}_{1100})+(\vec{a}_{0101}-\vec{a}_{1010})+(\vec{a}_{0110}-\vec{a}_{1001})$,\\ 
			
			$ \vec{m}_2=(\vec{a}_{0000}-\vec{a}_{1111})+(\vec{a}_{0001}-\vec{a}_{1110})+(\vec{a}_{0010}-\vec{a}_{1101})-(\vec{a}_{0100}-\vec{a}_{1011})-(\vec{a}_{0111}-\vec{a}_{1000})+(\vec{a}_{0011}-\vec{a}_{1100})-(\vec{a}_{0101}-\vec{a}_{1010})-(\vec{a}_{0110}-\vec{a}_{1001})$,\\
			
			$ \vec{m}_3=(\vec{a}_{0000}-\vec{a}_{1111})+(\vec{a}_{0001}-\vec{a}_{1110})-(\vec{a}_{0010}-\vec{a}_{1101})+(\vec{a}_{0100}-\vec{a}_{1011})-(\vec{a}_{0111}+\vec{a}_{1000})-(\vec{a}_{0011}-\vec{a}_{1100})+(\vec{a}_{0101}-\vec{a}_{1010})-(\vec{a}_{0110}-\vec{a}_{1001})$,\\
			
			$ \vec{m}_4=(\vec{a}_{0000}-\vec{a}_{1111})-(\vec{a}_{0001}-\vec{a}_{1110})+(\vec{a}_{0010}+\vec{a}_{1101})-(\vec{a}_{0100}-\vec{a}_{1011})-(\vec{a}_{0111}+\vec{a}_{1000})-(\vec{a}_{0011}-\vec{a}_{1100})-(\vec{a}_{0101}-\vec{a}_{1010})+(\vec{a}_{0110}-\vec{a}_{1001}).$  \label{eq:m}\\
			
			By observing the expressions for $\vec{m}_{y_{1}}$s, we can easily recognize that each Bloch vector must be a unit vector. In other words, the states should be pure unless it would lead to a decrement of the overall magnitude of $m_{y_{1}}$s. Along with this requirement, we can observe that among the sixteen unit vectors, eight appears with minus sign, the optimal magnitude of $\vec{m}_{y_{1}}$s would imply that all sixteen Bloch vectors would form eight antipodal pairs. Hence maximization of S$_{4}^{1}$ demands the two vectors in each parenthesis of Eq. \eqref{eq:m} are antipodal pairs, and hence we can write,\\
			
			$ \vec{m}_1=2(\vec{a}_{0000}+\vec{a}_{0001}+\vec{a}_{0010}+\vec{a}_{0100}+\vec{a}_{0111}+\vec{a}_{0011}+\vec{a}_{0101}+\vec{a}_{0110}$),  $ \vec{m}_2=2(\vec{a}_{0000}+\vec{a}_{0001}+\vec{a}_{0010}-\vec{a}_{0100}-\vec{a}_{0111}+\vec{a}_{0011}-\vec{a}_{0101}-\vec{a}_{0110}$), $ \vec{m}_3=2(\vec{a}_{0000}+\vec{a}_{0001}-\vec{a}_{0010}+\vec{a}_{0100}-\vec{a}_{0111}-\vec{a}_{0011}+\vec{a}_{0101}-\vec{a}_{0110}$), and $ \vec{m}_4=2(\vec{a}_{0000}-\vec{a}_{0001}+\vec{a}_{0010}+\vec{a}_{0100}-\vec{a}_{0111}-\vec{a}_{0011}-\vec{a}_{0101}+\vec{a}_{0110}$).\\
			In a compact form it can be written as $m_{y_{1}}=(-1)^{x^{i}_{y_1}}\vec{a}_{x^{i}}$ where $x_{y_{1}}^{i}$ is the $y_{1}^{th}$ bit of the $4$-bit input string $x^{i}\in \{0,1\}^{4}$  with first bit $0$, i.e., $x_{1}^{i}=0$ for all $i\in [8]$.  With this notation, the Eq. \eqref{eq:Succ41} can be re-written as
			\begin{equation}
				S_{4}^{1} =\frac{1}{2}+\frac{\alpha_{1}\beta_{1}}{16    }\sum_{x^{i}}\sum_{y_{1}=1}^{4} (-1)^{x_{y_{1}}^{i}}\vec{a}_{x^{i}}\cdot \hat{b}_{y_{1}}. \label{eq:Succ4111}
			\end{equation}
			
			Now using Eq. \eqref{eq:D} the reduced density operator after Bob$_{1}$'s measurement can be represented as,
			\begin{eqnarray}
				\rho^{1}_x  &&= \frac{1}{4}\sum_{y_{1},b} K_{b|y_{1}}\rho_{x}K_{b|y_{1}}								=\frac{\mathbb{I}+\vec{a}_{x}^{1}\cdot\sigma}{2},
			\end{eqnarray}
			
			where $\vec{a}_{x}^{1}$ is given by, 
			
			\begin{equation}
				\vec{a}^{1}_{x}= 2\left(\alpha_{1}^2-\beta_{1}^{2}\right)\vec{a}_{x} +\beta_{1}^{2}\sum_{y_{1}=1}^{4}\left( \hat{b}_{y_{1}}\cdot\vec{a}_{x}\right) \hat{b}_{y_{1}}. \label{eq:rho1}
			\end{equation}
			
			Using Eq. \eqref{eq:rho1} find the success probability of Bob$_{2}$ as, 
			\begin{equation}
				\begin{split}
					S_{4}^{2} &=\frac{1}{64}\sum_{x \in \{0,1\}^3}\sum_{y_{2}=1}^{4}Tr\Big[\rho_{x}^{1} B_{x_{y_{2}}|y_{2}}\Big]\\
					&=\frac{1}{2}+ \frac{1}{128}\Big[ 2\left(\alpha_{1}^2-\beta_{1}^{2}\right)\sum_{x\in \{0,1\}^{4}} \sum_{y_{2}=1}^{4}\left((-1)^{x_{y_2}}\vec{a}_{x}\cdot \hat{b}_{y_{2}}\right)+ \beta_{1}^{2}\sum_{x\in \{0,1\}^{4}} \sum_{y_{1},y_{2}=1}^{4}\left( (-1)^{x_{y_2}}\vec{a}_{x}\cdot\hat{b}_{y_{1}}\right)\left( \hat{b}_{y_{1}}\cdot \hat{b}_{y_{2}}\right)\Big]. \label{eq:Succ42a}
				\end{split}
			\end{equation}
			which we can re-write as
			\begin{equation}
				\begin{split}
					S_{4}^{2} =\frac{1}{2}+ \frac{1}{64}\Big[ 2\left(\alpha_{1}^2-\beta_{1}^{2}\right)\sum_{x^{i}} \sum_{y_{2}=1}^{4}\left((-1)^{x^{i}_{y_2}}\vec{a}_{x^{i}}\cdot \hat{b}_{y_{2}}\right)+ \beta_{1}^{2}\sum_{x^{i}} \sum_{y_{1},y_{2}=1}^{4}\left( (-1)^{x^{i}_{y_2}}\vec{a}_{x^{i}}\cdot\hat{b}_{y_{1}}\right)\left( \hat{b}_{y_{1}}\cdot \hat{b}_{y_{2}}\right)\Big]. \label{eq:Succ42b}
				\end{split}
			\end{equation}
			
			Let us now first maximize the first term in right hand side first. We consider a positive number $\gamma$ whose expectation value can be written as $\langle \gamma \rangle = \beta- \Delta$ , where $ \beta$ is a real number and $\Delta=\sum_{x^{i}} \sum_{y_{2}=1}^{4}(-1)^{x^{i}_{y_1}}\vec{a}_{x^{i}}\cdot \hat{b}_{y_{2}}$. Furthermore, we consider a set of positive vectors $\vec{L}_{i}$ which is polynomial functions of $\vec{a}_{x^{i}}$ and $b_{y_{2}}$ so that $\gamma$ acquires the form,
			\begin{equation}
				\gamma=\sum_{i=1}^{8}\frac{w_{i}}{2}(\vec{L}_{i})^{2}. \label{eq:gamma}
			\end{equation}
			
			Suitable vectors $\vec{L}_{i}$ can be chosen for function given in equation \eqref{eq:gamma} as,
			\begin{equation}
				\vec{L}_{i}=\frac{1}{w_{i}}\sum_{y=1}^{n} (-1)^{x_{y}^{i}}\hat{b}_{y} -\vec{a}_{x^i}, \label{eq:L}
			\end{equation}
			
			where $w_{i}$ is any positive semi definite function of $b_{y_{2}}$. Now substituting Eq. \eqref{eq:L} into Eq. \eqref{eq:gamma} and noting that $(\hat{b}_{y_{2}})^{2}=(\vec{a}_{x^i})^{2}=1$, we get,
			\begin{equation}
				\langle\gamma\rangle=\sum_{i=1}^{8}\Big[\frac{1}{2w_{i}}\Big(\sum_{y=1}^{4} (-1)^{x_{y_{2}}^{i}}\hat{b}_{y_{2}}\Big)^2 + \frac{w_{i}}{2}\Big] - \Delta .\label{eq:gamma2}
			\end{equation}
			We can now conveniently put $w_{i}=||\sum_{y=1}^{4} (-1)^{x_{y_{2}}^{i}}b_{y_{2}}||$ in Eq.\eqref{eq:gamma2} to finally get, $\Delta=\sum_{i=1}^{8}w_{i}- \langle\gamma\rangle$.		Since $\langle\gamma\rangle\geq0$ the maximum quantum value of $\Delta$  can  be written as 
			\begin{equation}
				(\Delta)_{max}=\max_{\hat{b}_{y_{2}}}\Bigg(\sum_{i=1}^{8}w_{i}\Bigg).
			\end{equation} 
			
			Since maximum value of $\Delta$ demands $\langle\gamma_{n}\rangle=0$ which further implies that $\vec{L}_{i}=0$ and consequently we have the input Bloch vectors $		\vec{a}_{x^i}=\frac{		\sum_{y=1}^{n} (-1)^{x_{y}^{i}}\hat{b}_{y}}{w_{i}}$. 
			
			Using the concavity inequality $\sum_{i=1}^{n}w_{i}\leq\sqrt{n\sum_{i=1}^{n}(w_{i})^2}$, 	and applying it two times we have 
			\begin{eqnarray}
				(\Delta)_{max}&\leq \sqrt{ 4(w_{1}^2+w_{2}^2+w_{3}^2+w_{4}^2)} +\sqrt{ 4(w_{5}^2+w_{6}^2+w_{7}^2+w_{8}^2)}\label{ww}\\
				&=4\sqrt{4+ 2\hat{b}_{3}\cdot\hat{b}_{4} } +4\sqrt{4- 2\hat{b}_{3}\cdot\hat{b}_{4} }. \nonumber
			\end{eqnarray}
			In the first line of Eq. (\ref{ww}), the equality holds when $w_{1}=w_{2}=w_{3}=w_{4}$ and same for the third term. This provides $\hat{b}_{1}\cdot\hat{b}_{2}=\hat{b}_{1}\cdot\hat{b}_{3}=\hat{b}_{2}\cdot\hat{b}_{3}=0$. Then, in qubit system it is not possible to obtain $\hat{b}_{3}\cdot\hat{b}_{4}=0$. The maximum value of Eq. (\ref{ww}) can be obtained when $\hat{b}_{3}\cdot\hat{b}_{4}=1$. We then have,
			
			\begin{eqnarray}
				\label{www}
				(\Delta)_{max}=4\left(\sqrt{2} +\sqrt{6}\right).
			\end{eqnarray}
			We can then chose $\hat{b}_{1}=\hat{x}$, $\hat{b}_{2}=\hat{y}$, $\hat{b}_{3}=\hat{z}$ and $\hat{b}_{4}=\hat{z}$. This in turn provides from the optimization condition $\hat{L}_{i}=0$ that $m_{y_{1}}=(-1)^{x_{y}^{i}}\vec{a}_{x^i}=\hat{b}_{y_{2}}$. Thus the input Bloch vectors $\vec{a}_{x^i}$ can be written as 		
			
			\begin{eqnarray}
				\vec{a}_{0000} &&= \frac{\hat{x}+\hat{y}}{\sqrt{2}}, \hspace{0.8cm}
				\vec{a}_{0001}=\frac{\hat{x}+\hat{y}+2\hat{z}}{\sqrt{6}},
				\hspace{0.8cm}
				\vec{a}_{0010}= \frac{\hat{x}+\hat{y}-2\hat{z}}{\sqrt{6}}, \hspace{0.8cm}
				\vec{a}_{0100}=\frac{\hat{x}-\hat{y}}{\sqrt{2}},\\
				\nonumber
				\vec{a}_{1000}&=& \frac{-\hat{x}+\hat{y}}{\sqrt{2}}, \hspace{0.8cm}
				\vec{a}_{0011}= \frac{\hat{x}+\hat{y}}{\sqrt{2}}, \hspace{0.8cm}
				\vec{a}_{0101}= \frac{\hat{x}-\hat{y}+2\hat{z}}{\sqrt{6}},  \hspace{0.8cm}
				\vec{a}_{0110}= \frac{\hat{x}-\hat{y}-2\hat{z}}{\sqrt{6}},
			\end{eqnarray}
			It can be easily checked that the second term in Eq. (\ref{eq:Succ42b}) is maximized when $\hat{b}_{y_2}=\hat{b}_{y_1}$ when $y_{1}=y_{2}$.  By putting the  values of $\Delta$, $\alpha$ and $\beta$ we have 
			\begin{equation}
				max(S_{4}^{2}) = \Omega_{2} =\frac{1}{2}+ \lambda_{2}\Bigg( \frac{(\sqrt{2}+\sqrt{6})}{64}\Bigg)\big(1+3\sqrt{1-\lambda_{1}^2}\big) \label{eq:Succ421}
			\end{equation}
			
			and consequently 
			
			\begin{equation}
				max(S_{4}^{1}) = \Omega_{1} =\frac{1}{2}+ \lambda_{1}\Bigg( \frac{(\sqrt{2}+\sqrt{6})}{16}\Bigg).\label{eq:Succ411}
			\end{equation}

			Considering the situation when Bob$_{1}$ and Bob$_{2}$ implement their unsharp measurements so that the quantum advantage to Bob$_{3}$ persists, we can calculate the minimum value of the unsharpness parameters from Eqs. \eqref{eq:Succ411} and \eqref{eq:Succ421} as, $\lambda_{1}^{min}=0.776$, $\lambda_{2}^{min}=1.074$. It is clear now that as $\lambda_{2}^{min}\geq 1$ is not a legitimate value, Bob$_{2}$ does not get any quantum advantage. But we can change the scenario by putting the restriction of parity-obliviousness to see whether in 4-bit RAC quantum advantage can be extended to Bob$_{2}$.
			
			\subsection{Maximum success probability for  parity-oblivious RAC for qubit system}

			Let us now discuss the case when the same game is constrained by parity-obliviousness constrains. In this case the classical bound of the game becomes the preparation non contextual bound and is equal to, $S_{4}=\frac{1}{2}(1+\frac{1}{4})\approx 0.625$. We recall the parity set $ \mathbb{P}_n= \{x|x \in \{0,1\}^n,\sum_{r} x_{r} \geq 2\} $ with $r\in \{1,2,...,n\}$. For any $s \in \mathbb{P}_{n}$, no information about $s\cdot x = \oplus_{r} s_{r}x_{r}$ (s-parity) is to be transmitted to Bob, where $\oplus$ is sum modulo $ 2 $. We then have s-parity-0 and s parity-1 sets. For 4-bit case the cardinality of parity set is eleven. Among them only four elements, 1011, 0111, 1110 and 0111, provide nontrivial constraints on the inputs are respectively given by

			\begin{eqnarray}  &&R_{1011}=(\vec{a}_{0000}-\vec{a}_{0001}-\vec{a}_{0010}+\vec{a}_{0100}-\vec{a}_{1000}+\vec{a}_{0011}-\vec{a}_{0101}-\vec{a}_{0110}) =0  \nonumber\\ &&R_{0111}= (\vec{a}_{0000}-\vec{a}_{0001}-\vec{a}_{0010}-\vec{a}_{0100}+\vec{a}_{1000}+\vec{a}_{0011}+\vec{a}_{0101}+\vec{a}_{0110}) =0\nonumber\\
				&&R_{1110}=(\vec{a}_{0000}+\vec{a}_{0001}-\vec{a}_{0010}-\vec{a}_{0100}-\vec{a}_{1000}-\vec{a}_{0011}-\vec{a}_{0101}+\vec{a}_{0110})=0 \nonumber \\ &&R_{1101}= (\vec{a}_{0000}-\vec{a}_{0001}+\vec{a}_{0010}-\vec{a}_{0100}-\vec{a}_{1000}-\vec{a}_{0011}+\vec{a}_{0101}-\vec{a}_{0110})=0 . \label{eq:R}
			\end{eqnarray}
			When the parity-oblivious restriction is imposed on Alice's inputs then we get $\Delta_{max}=8\sqrt{2}$. This in turn provides the maximum success probability of Bob$_{2}$, 
			
			\begin{equation}
				max(S_{4}^{2}) = \Omega_{2} =\frac{1}{2}+ \lambda_{2}\Bigg( \frac{\big(1+3\sqrt{1-\lambda_{1}^2}\big)}{16\sqrt{2}}\Bigg), \label{eq:Succ42a1}
			\end{equation}
			
			which also fixes the maximum success probability of Bob$_{1}$ in 4-bit parity-oblivious RAC is given by,
			
			\begin{equation}
				max(S_{4}^{1}) = \Omega_{1} =\frac{1}{2}+ \lambda_{1}\Bigg( \frac{1}{4\sqrt{2}}\Bigg).\label{eq:Succ41a1}
			\end{equation}

			The respective minimum values of sharpness parameter required to get quantum advantage for Bob$_{1}$ and Bob$_{2}$ are $\lambda_{1}^{min}=0.707$, $\lambda_{2}^{min}=0.906$ . Thus both Bob$_{1}$ and Bob$_{2}$ both get the quantum advantage. We then extend our calculation to Bob$_{3}$ to check whether he gets any quantum advantage or not. As a continuation to the previous section we can write the reduced state after Bob$_{2}$'s measurement will be,
			
			\begin{eqnarray}
				\rho^{2}_x = \frac{1}{16} \Bigg\{ 8\mathbb{I} + 32\alpha_{1}^2 \alpha_{2}^2(\vec{a}_{x} \cdot\sigma) + 8\alpha_{2}^2\beta_{1}^2\sum_{y_{1}=1}^{4}B_{y_{1}}(\vec{a}_{x}\cdot \sigma) B_{y_{1}} +8\alpha_{1}^2 \beta_{2}^2\sum_{y_{2}=1}^{4}B_{y_{2}}(\vec{a}_{x}\cdot \sigma) B_{y_{2}}+ 2\beta_{1}^2\beta_{2}^2\sum_{y_{1},y_{2}=1}^{4}B_{y_{2}}B_{y_{1}}(\vec{a}_{x}\cdot \sigma)B_{y_{1}} B_{y_{2}}\Bigg\}.
			\end{eqnarray}
			
			After some simplification we can represent above equation as $\rho^{2}_x = \frac{\mathbb{I}+\vec{a}_{x}^{2}\cdot\sigma}{2}$ where the Bloch vector $\vec{a}_{x}^{2}$ can be written as,
			
			\begin{eqnarray}
				\vec{a}^{2}_{x}= 4\left(\alpha_{1}^2-\beta_{1}^{2}\right)\left(\alpha_{2}^2-\beta_{2}^{2}\right)\vec{a}_{x}+\beta_{1}^{2}\left(\alpha_{2}^2-\beta_{2}^{2}\right) \sum_{y_{1}=1}^{4}\left( \hat{b}_{y_{1}}\cdot\vec{a}_{x}\right) \hat{b}_{y_{1}} +\beta_{2}^{2}\left(\alpha_{1}^2-\beta_{1}^{2}\right) \sum_{y_{2}=1}^{4}\left( \hat{b}_{y_{2}}\cdot\vec{a}_{x}\right) \hat{b}_{y_{2}}+\beta_{1}^2\beta_{2}^{2}\sum_{y_{1},y_{1}=1}^{4}\left( \hat{b}_{y_{1}}\cdot\vec{a}_{x}\right)\left( \hat{b}_{y_{1}}\cdot \hat{b}_{y_{2}}\right) \hat{b}_{y_{2}} \label{eq:4222}. \nonumber\\
			\end{eqnarray}
			
			The success probability for Bob$_{3}$ is
			
			\begin{eqnarray}
				S_{4}^{3} &&= \frac{1}{2} +\frac{1}{128}\Bigg[ 4\left(\alpha_{1}^2-\beta_{1}^{2}\right)\left(\alpha_{2}^2-\beta_{2}^{2}\right)\sum_{x^{i}} \sum_{y_{2}=1}^{4}\left((-1)^{x^{i}_{y_1}}\vec{a}_{x^{i}}\cdot \hat{b}_{y_{2}}\right)+\beta_{1}^{2}\left(\alpha_{2}^2-\beta_{2}^{2}\right) \sum_{x^{i}} \sum_{y_{1},y_{2}=1}^{4}\left( (-1)^{x_{y_1}}\vec{a}_{x^{i}}\cdot\hat{b}_{y_{1}}\right)\left( \hat{b}_{y_{1}}\cdot \hat{b}_{y_{2}}\right) \nonumber \\  &&+\beta_{2}^{2}\left(\alpha_{1}^2-\beta_{1}^{2}\right) \sum_{x^{i}} \sum_{y_{2},y_{3}=1}^{4}\left((-1)^{x^{i}_{y_1}}\vec{a}_{x^{i}}\cdot \hat{b}_{y_{2}}\right)\left( \hat{b}_{y_{2}}\cdot \hat{b}_{y_{3}}\right)+\beta_{1}^2\beta_{2}^{2}\sum_{x^{i}} \sum_{y_{1},y_{2},y_{3}=1}^{4}\left((-1)^{x^{i}_{y_1}}\vec{a}_{x^{i}}\cdot \hat{b}_{y_{2}}\right)\left( \hat{b}_{y_{1}}\cdot \hat{b}_{y_{2}}\right)\left( \hat{b}_{y_{2}} \cdot \hat{b}_{y_{3}}\right) \Bigg] . \label{eq:Succ43a}
			\end{eqnarray}

			We finally obtain the maximum success probability for Bob$_{3}$ as,
			
			\begin{equation}
				max(S_{4}^{3}) =\Omega_{3} =\frac{1}{2} + \Big(\frac{ \sqrt{2}}{128}\Big)\prod_{k^{'}=1}^{2}(1+3\sqrt{1-\lambda_{k^{'}}^2}). \label{eq:succ441a}
			\end{equation}

			The minimum sharpness parameter that Bob$_3$'s instrument has to be $\lambda_{3}^{min}=1.56\geq1$. Since this value is not legitimate we have no advantage for Bob$_{3}$.

			\subsection{Optimal success probability in parity-oblivious RAC for two-qubit system }
			We provide sketch of the argument when Alice's input states are two-qubit system. From the entanglement assisted parity-oblivious RAC in \cite{ghorai18}, in our prepare-measure scenario we find that  to obtain the optimal quantum success probability the input states has to be 
			
			\begin{align}
				\rho_{x}=\frac{1}{4}\left(\mathbb{I}\otimes\mathbb{I} +\frac{(-1)^{x_{y}^{i}}B_{y}}{2}\right)
			\end{align}
			where $B_{y}$s are mutually anticommuting observables in two-qubit system. One of such choices are $B_1=\sigma_{x}\otimes\sigma_{x}$, $B_2=\sigma_{x}\otimes\sigma_{y}$, $B_3=\sigma_{x}\otimes\sigma_{z}$ and $B_4=\sigma_{y}\otimes\mathbb{I}$. Imortantly, the sixteen  two-qubit input states corresponding to $x\in \{0,1\}^{4}$ satisfy the four nontrivial parity-oblivious conditions. The optimal quantum success probability for Bob$_{1}$ can be obtained as,
			
			\begin{equation}
				max(S_{4}^{1}) =\xi_{4}^{1} =\frac{1}{2}+  \frac{\lambda_{1}}{4}\label{eq:Succtwo2}
			\end{equation}
			and for Bob$_{2}$, 
			\begin{equation}
				max(S_{4}^{2}) =\xi_{4}^{2} =\frac{1}{2}+ \lambda_{2}\Bigg( \frac{1}{16}\Bigg)\big(1+3\sqrt{1-\lambda_{1}^2}\big). \label{eq:Succtwo1}
			\end{equation}
			
			The general form of optimal quantum success probability for k$^{th}$ Bob can be written as,

			\begin{equation}
				max(S_{4}^{k}) =\xi_{4}^{k} =\frac{1}{2}+ \lambda_{2}\Bigg( \frac{1}{4^{k}}\Bigg)\prod_{k^{'}=1}^{k-1}(1+3\sqrt{1-\lambda_{k^{'}}^2}). \label{eq:Succtwok}
			\end{equation}
			
			The preparation non-contextual bound on $4$-bit parity-oblivious RAC is $S_{4}=\frac{1}{2}(1+\frac{1}{4})\approx 0.625$. In this case we get the minimum value of unsharpness parameters of five sequential Bobs are  $\lambda_{1}^{min}=0.5$, $\lambda_{2}^{min}=0.556$, $\lambda_{3}^{min}=0.679$, $\lambda_{4}^{min}=0.795$ and $\lambda_{5}^{min}=1.12\geq1$. Hence, if Alice's encoding quantum states are in two-qubit system, at most four Bobs can get quantum advantage. Similar trade-off relation can be demonstrated by following the prescription in $3$-bit parity-oblivious RAC.			
		\end{widetext}

	\end{document}